%% file: paper.tex
\documentclass[conference,10pt,letterpaper]{IEEEtran}
\IEEEoverridecommandlockouts

\usepackage{amsmath,amssymb,mathtools}
\usepackage{booktabs,multirow,array,tabularx}
\usepackage{graphicx}
\usepackage{tikz}
\usetikzlibrary{arrows.meta,calc,fit,positioning}
\usepackage{enumitem}
\usepackage{placeins}
\usepackage{capt-of}
\usepackage{microtype}
\usepackage{xurl}
\usepackage{cite}
\usepackage[hidelinks]{hyperref}
\usepackage{cleveref}
\usepackage{xspace}

\newcolumntype{L}[1]{>{\raggedright\arraybackslash}p{#1}}
\newcolumntype{C}[1]{>{\centering\arraybackslash}p{#1}}
\newcolumntype{Y}{>{\raggedright\arraybackslash}X}

\newcommand{\Req}{\ensuremath{\mathsf{Req}}}
\newcommand{\Norm}{\ensuremath{\mathsf{Norm}}}
\newcommand{\CheckPlan}{\ensuremath{\mathsf{CheckPlan}}}
\newcommand{\VerifyEvidence}{\ensuremath{\mathsf{VerifyEvidence}}}

\newcommand{\SubObj}{\ensuremath{\mathsf{SubObj}}}
\newcommand{\Cov}{\ensuremath{\mathsf{Cov}}}
\newcommand{\SubmittedValid}{\ensuremath{\mathsf{SubmittedValid}}}
\newcommand{\CompleteRealization}{\ensuremath{\mathsf{CompleteRealization}}}
\newcommand{\SOVEAccept}{\ensuremath{\mathsf{SOVEAccept}}}
\newcommand{\OpRep}{\ensuremath{\mathsf{OpRep}}}
\newcommand{\StructOK}{\ensuremath{\mathsf{StructOK}}}
\newcommand{\CompAccept}{\ensuremath{\mathsf{ComponentAccept}}}
\newcommand{\StructureAccept}{\ensuremath{\mathsf{StructureAccept}}}
\newcommand{\RealizationGap}{\ensuremath{\mathsf{RealizationGap}}}
\newcommand{\AdmissionSOVE}{\ensuremath{\mathsf{AdmissionSOVE}}}
\newcommand{\FalseRelease}{\ensuremath{\mathsf{FalseRelease}}}

\newcommand{\EntryComplete}{\ensuremath{\mathsf{EntryComplete}}}
\newcommand{\MaterializationComplete}{\ensuremath{\mathsf{MaterializationComplete}}}
\newcommand{\ObjectBind}{\ensuremath{\mathsf{ObjectBind}}}
\newcommand{\FinalizeGuard}{\ensuremath{\mathsf{FinalizeGuard}}}
\newcommand{\FinalAccept}{\ensuremath{\mathsf{FinalAccept}}}
\newcommand{\Consumable}{\ensuremath{\mathsf{Consumable}}}
\newcommand{\Reach}{\ensuremath{\mathsf{Reach}}}
\newcommand{\ExactRealizationTheorem}{Theorem~1}
\newcommand{\SemanticCompositionProposition}{Proposition~1}
\newcommand{\SourceTransferClaim}{Claim~1}
\newcommand{\Pzero}{\textsc{RelationIR-Core v1}\xspace}
\newcommand{\system}{H-CRSPV\xspace}
\newcommand{\sove}{SOVE\xspace}

\newcommand{\FinalChainAttacks}{\VtwelvePending{chain attacks}}

\newcommand{\FinalWrongDigestResult}{\VtwelvePending{wrong-digest result}}

\newcommand{\FinalRealizationEntries}{\VtwelvePending{entries}}
\newcommand{\FinalMaterializationGroups}{\VtwelvePending{groups}}
\newcommand{\FinalActualEvidenceObjects}{\VtwelvePending{objects}}
\newcommand{\FinalCertifiedLocalEntries}{\VtwelvePending{local entries}}

\newcommand{\FinalEOneLegal}{\VtwelvePending{legal partitions}}
\newcommand{\FinalEOneInvalid}{\VtwelvePending{invalid partitions}}

\newcommand{\FinalETwoAttacks}{\VtwelvePending{baseline attacks}}

\newcommand{\FinalAlternateDigests}{\VtwelvePending{alternate digests}}
\newcommand{\FinalQwenValidatorSeconds}{\VtwelvePending{validator seconds}}
\newcommand{\FinalQwenReplaySeconds}{\VtwelvePending{replay seconds}}
\input{v12_1_results_macros.tex}
\newcommand{\VThirteenQwenModel}{}

\newcommand{\VThirteenLegalStates}{}

\newcommand{\VThirteenConvergedCases}{}
\newcommand{\VThirteenChallengeWindow}{}

\newcommand{\VThirteenFreshReplayBytes}{}
\newcommand{\VThirteenFreshEvidenceBytes}{}
\newcommand{\VThirteenFreshByteRatio}{}
\newcommand{\VThirteenProviderPlaneBytes}{}

\newcommand{\VThirteenFreshInclusiveRatio}{}
\newcommand{\VThirteenFreshReusableBytes}{}
\newcommand{\VThirteenFreshCacheHitBytes}{}
\newcommand{\VThirteenFreshNewBytes}{}
\newcommand{\VThirteenFreshPerValidatorNewBytes}{}

\newcommand{\VThirteenWarmReplayIncrementalBytes}{}
\newcommand{\VThirteenWarmEvidenceIncrementalBytes}{}

\newcommand{\VThirteenBehaviorSamples}{}

\newcommand{\VThirteenProviderProcesses}{}
\newcommand{\VThirteenIsolationReceipts}{}
\newcommand{\VThirteenPrivateCanarySuccesses}{}
\newcommand{\VThirteenPrivateKeyMounts}{}
\newcommand{\VThirteenSignerSockets}{}
\input{v13_results_macros.tex}

\setlist[itemize]{leftmargin=*,nosep}
\setlist[enumerate]{leftmargin=*,nosep}
\newcommand{\TableFont}{\footnotesize}
\newcommand{\State}[1]{\textsc{#1}}

\title{H-CRSPV: Preventing Semantic Omission\\
in Late-Bound Large Language Model Releases}

\input{authors.tex}

\begin{document}
\maketitle
\pagestyle{plain}

\begin{abstract}
Large-language-model release pipelines increasingly combine commitments,
signatures, provenance records, and heterogeneous verification backends. Yet
validating every submitted object does not establish that a release realizes
every requirement of its registered transformation. An untrusted realization
proposer may omit a required relation, propose an unauthorized evidence-sharing
assignment, or bind valid evidence to the wrong object. We term this
verification-boundary failure \emph{Semantic Omission under Valid Evidence}
(\sove).

\emph{Hybrid Cryptographic Relation-based Semantic Plan Verification}
(H-CRSPV) is a replicated admission layer that enforces required-set
completeness before release consumption. Before evidence selection, an
authorized registration entity commits an authenticated authority record. The
record fixes a closed, typed RelationIR specification, output identities,
admissible evidence policies, and lifecycle rules; it is not an online planning
role. The untrusted realization proposer later selects evidence methods and
sharing assignments according to current availability. Validators independently
derive the required-obligation multiset, check exact entry-occurrence coverage,
validate proposal-induced evidence sharing, and bind admissible groups
one-to-one to keeper-resolved objects. Backend-specific verification runs only
after these structural checks succeed. Evidence remains provisional until the
challenge window closes and an atomic finalizer activates the release.

The analysis establishes structural exactness and conditional semantic
guarantees under explicit assumptions. Across 36 omission artifacts,
submitted-object validation accepts all 36 because every submitted object
passes its backend-specific verifier; H-CRSPV rejects all 36 while accepting
all six honest releases. The prototype also validates six restricted
source-to-RelationIR bindings and rejects all 60 tested mutations. In a
continuous four-validator \VThirteenQwenModel{} workflow, the same authority
record remains fixed across three legal releases with different
post-registration availability states. No re-registration is required, and all
four validators agree in rejecting four targeted attacks. Per-object validity
therefore does not establish complete, correctly bound, and finalized release
evidence.
\end{abstract}

\section{Introduction}
\label{sec:intro}

Training a frontier-scale large language model is a substantial capital
investment. Recent estimates place hardware and staff expenses for
GPT-4- and Gemini-class systems in the tens of millions of dollars and project
that the largest training runs could exceed one billion dollars by 2027
\cite{cottier2024rising}. A growing market prices model access through hosted
inference APIs, while open-weight ecosystems support local reuse and derivative
releases~\cite{demirer2025intelligence}. Model weights and their derivatives
are therefore economically valuable, versioned production artifacts.

That value does not stop at a base checkpoint.
Low-rank adapters, sparse updates, merged checkpoints, quantized releases, and
sharded model files may be produced, certified, published, and consumed by
different parties. Their release pipelines increasingly combine commitments,
signatures, provenance records, and fraud proofs from several evidence
backends. Those mechanisms establish whether a submitted opening, equation,
trace step, or signature is valid. They do not by themselves establish that
the release contains every statement required by the declared transformation.

This gap appears when evidence realization is selected after task registration.
Consider a fixed-point merge of a base checkpoint and two LoRA adapters. The
release must bind affine composition, signed ranges, rounding, adapter-to-output
reindexing, and the consumer-visible output. An untrusted proposer can submit a
valid KZG relation for the affine part, a valid BLS aggregate over the local
relations it chose, and valid availability metadata, yet omit a required range
or reindex relation or point a proof to a stale intermediate object. Every
visible cryptographic check succeeds for its submitted statement, but the
release has not realized the registered transformation.

We term this interface failure \emph{Semantic Omission under Valid Evidence}
(\sove). The vulnerable abstraction quantifies over submitted objects; the
required security judgment quantifies over the registered obligation set and
its concrete realization. Mixed-backend pipelines make the distinction
consequential because a single release may distribute obligations across
direct checks, commitment proofs, and committee-certified local relations.

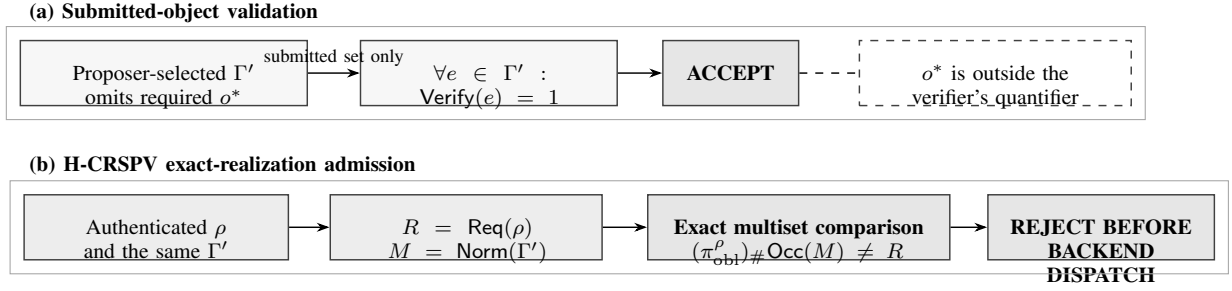
\begin{figure*}[t]
  \centering
  \input{figures/sove_attack_tikz.tex}
  \caption{Submitted-object validation quantifies only over evidence in
  $\Gamma$ and therefore cannot detect an omitted requirement. H-CRSPV
  independently reconstructs $\Req(\rho)$ and rejects an entry-multiset
  mismatch before backend dispatch.}
  \label{fig:sove}
\end{figure*}

H-CRSPV makes this completeness boundary a replicated admission decision
(Fig.~\ref{fig:sove}). At registration, an authenticated authority fixes the
program, output identities, required obligations, admissible backends, batching
policy, and challenge policy. Evidence and availability may change later, so
the proposer chooses how to realize each obligation and which entries propose
to share evidence. Validators reconstruct that proposed grouping from complete
keys and accept it only when the registered batching policy permits every
group. Each accepted group then binds to exactly one concrete object resolved
from replicated state. Native verification runs over those objects, and only a
finalized release becomes active for downstream use.

\paragraph{Novelty and contributions}
\textbf{SOVE as a verification-boundary failure.} Late-bound evidence
orchestration exposes a security-critical boundary: every submitted object may
pass backend verification while the realization still fails to cover the
obligation multiset derived from the registered transformation. SOVE formalizes
this gap between per-object validity and complete release realization.
\textbf{Authority-owned exact-realization admission.} A closed, typed
RelationIR profile and deterministic constructor derive a canonical,
multiplicity-preserving obligation multiset independently of proposer evidence.
H-CRSPV carries this requirement set through policy-checked evidence sharing,
keeper-resolved object binding, and lifecycle-gated admission.
\textbf{Replicated end-to-end enforcement.} A Python, Rust, and
Cosmos/CometBFT prototype integrates restricted PEFT/Safetensors bindings with
native KZG and BLS evidence and evaluates omission, illegal sharing,
object-binding failure, dynamic late binding, lifecycle safety, replica
agreement, and fair completeness baselines.

\paragraph{Key results}
The restricted frontend binds six source programs and rejects 60/60 source
mutations. Submitted-object validation accepts 36/36 loadable omission
artifacts; H-CRSPV and three fair complete baselines reject all 36 while
accepting all six honest runs.
For one provenance-bound \VThirteenQwenModel{} artifact under a fixed
authority, all three releases finalize, and all four validators reject all
four attacks. H-CRSPV adds concrete-object coverage and replicated activation.

Our claims cover the closed \Pzero profile and the evaluated declarative
PEFT/Safetensors subset; backend soundness, cross-commitment co-reference, and
optimistic finalization retain the explicit assumptions stated in
Sec.~\ref{sec:threat}.

\section{Background and Problem Statement}
\label{sec:sove}

\subsection{Late-Bound Evidence Pipelines}
A task creator or governance registers an immutable authority that fixes the
execution context, profile and typed program, policy and catalogs, tensor and
commitment descriptors, consumer output, production setup, certifier epoch and
quorum, and workflow identity. Evidence may repeat those fields for binding but
cannot replace consensus task state. After registration, an untrusted proposer
selects realizations, batching, and witnesses. A submitted-object stack then
checks the normalized relations it receives. That check is necessary, but its
quantifier ranges over a proposer-selected set rather than every requirement
derived from the authority; Sec.~\ref{sec:design} formalizes the distinction.

The boundary arises when semantics induce finite requirements, a later actor
chooses their concrete realizations, and component verifiers quantify only over
submitted objects. In a PEFT/LoRA workflow, base and adapter tensors pass
through merge, sharding, and publication; late adapter-shard or backend
selection can omit a reindex, range, or rounding obligation. In a quantized
release, calibration produces scales and tensor bundles; a late manifest can
omit a tensor or cross-bind its scale. In a zkML marketplace, individually
valid subgraph proofs can fail to cover the registered graph
~\cite{hu2022lora,lycklama2024artemis,sun2023zkdl}.

Formally, an authenticated authority $\rho$ determines the finite required
obligation multiset
\begin{equation}
R_\rho:=\Req(\rho)\in\mathsf{MSet}(\mathcal U_\rho).
\label{eq:abstract-required}
\end{equation}
At release time, an untrusted proposer submits $q$, containing a realization
proposal, evidence payloads or authenticated object references, and a candidate
output. The abstract judgment evaluates $q$ against the authenticated
admission-time keeper snapshot. Let $\SubObj(q)$ be the distinct set of objects
resolved for that release. Submitted-object validation is
\begin{equation}
\SubmittedValid_{\rho,\ell}(q):=
 \bigwedge_{e\in\SubObj(q)}
 \mathsf{Verify}^{\mathrm{obj}}_{\rho,b_e}(\sigma_e,e)=1.
\label{eq:submitted-valid}
\end{equation}
This predicate may contain strong syntax, cryptographic, and context checks; its
quantifier nevertheless ranges only over objects selected by $q$.

\subsection{Semantic Omission and the Completeness Boundary}
Complete realization is independent of any backend data structure. For each
$e\in\SubObj(q)$, the partial function $\Cov_{\rho,\ell}(q,e)$ denotes the exact
obligation multiset that $e$ is structurally authorized to realize in the
current task, release, candidate-output, and binding context. It is undefined
for cross-bound or orphan objects, unauthorized sharing, duplicate object keys,
or any object with no authority-scoped role. This is a structural association,
not a semantic-truth claim. Define
\begin{equation}\small
\begin{aligned}
\mathsf{CovDefined}_{\rho,\ell}(q):={}&
 \forall e\in\SubObj(q):
 \Cov_{\rho,\ell}(q,e)\mathord\downarrow,\\
\CompleteRealization_{\rho,\ell}(q):={}&
 \mathsf{CovDefined}_{\rho,\ell}(q)\\[-0.25ex]
&\land\biguplus_{e\in\SubObj(q)}
 \Cov_{\rho,\ell}(q,e)=R_\rho.
\end{aligned}
\label{eq:complete-realization}
\end{equation}
The two clauses reject missing or duplicated occurrences, unauthorized overlap,
cross-context association, and objects with no registered role. They neither
require one object per obligation nor assert semantic truth or finality.

For an interface $I$, abstract SOVE acceptance and consumer safety are
\begin{equation}
\begin{split}
\SOVEAccept^I_{\rho,\ell}(q):={}&
 \mathsf{Accept}^I_{\rho,\ell}(q)\\[-0.25ex]
&\land\SubmittedValid_{\rho,\ell}(q)\\[-0.25ex]
&\land\neg\CompleteRealization_{\rho,\ell}(q),\\
\Consumable^I_{\rho,\ell}(q,s)\Longrightarrow{}&
 \SubmittedValid_{\rho,\ell}(q)\\[-0.25ex]
&\land\CompleteRealization_{\rho,\ell}(q).
\end{split}
\label{eq:abstract-sove}
\end{equation}
The first predicate distinguishes SOVE from backend unsoundness; the second
separates evidence validity and exact realization from lifecycle finality.
$\Cov$ is not an added oracle. Section~\ref{sec:design} instantiates it through
proposal-induced groups that pass $\mathsf{GroupOK}$ and the keeper-resolved
$\ObjectBind$ bijection; Lemma~\ref{lem:bridge} proves the correspondence.

A faithful complete circuit, deterministic replay, or trusted complete planner
already closes the omission gap. The missing judgment appears only when an
adversary selects the statement set and the verifier has neither a trusted
complete set, a reconstruction procedure, nor a proof of set equality.
Composition preserves the validity of received objects; it cannot reveal an
object never requested.

A fixed semantic inspector can close identifier-set omission by comparing the
authority-derived and submitted requirement IDs. H-CRSPV carries a stronger
field-level check through registered-policy validation of proposer grouping,
concrete object binding, native evidence, workflow binding, and finalization.
Section~\ref{sec:design} refines these predicates operationally;
Table~\ref{tab:boundary-ownership}
summarizes the resulting ownership boundary.

\begin{table*}[t]
\centering
\caption{Responsibility boundaries of the evaluated completeness mechanisms.}
\label{tab:boundary-ownership}
\TableFont
\begin{tabularx}{\textwidth}{@{}L{0.17\textwidth}YYY@{}}
\toprule
Boundary decision & Submitted-object validation & Fixed semantic inspector & \textbf{H-CRSPV keeper} \\
\midrule
Required-set owner & Proposer manifest & Fixed policy & Consensus authority \\
Entry-set check & None & Requirement-ID equality & Exact normalized multiset \\
Grouping decision & Backend-local & Not checked & Registered-policy validation of proposed quotient \\
Object binding & Submitted handle & Inspector-visible input & Keeper-resolved bijection \\
Native evidence & Per-backend contracts & Inspector predicate & Context-bound contracts \\
Consumability & External to interface & External to interface & Replicated challenge and atomic finalization \\
\bottomrule
\end{tabularx}
\end{table*}

\subsection{Why Late Binding Matters}
Late binding matters when proof availability, caches, quorum or DA state, or
cost changes after registration. The authority fixes \emph{what}; the untrusted
proposer chooses only \emph{how}. A trusted complete planner or deterministic
replay can enforce the same requirement coverage. H-CRSPV is designed for the
case in which realization crosses a trust boundary and consumption requires a
shared replicated decision.

\section{Threat Model and Security Goals}
\label{sec:threat}

\subsection{Parties and Adversary}
Before any manifest is accepted, the keeper loads the task Verification Authority from consensus task state. It authenticates $H(P)$, $H(\Sigma)$, $H(C)$, $H(D)$, $H(I)$, every input/output descriptor and root mapping, the consumer-visible output identity, SRS/backend allowlists, the four-member certifier set, epoch and quorum, and the workflow layout/key. Updating any of these fields creates a new authority version; a proposer cannot create, replace, or downgrade the authority for an existing task by repeating different values in an evidence message. The authority may be registered by a release owner or governance, but its semantic adequacy remains explicitly in the TCB.

Authentication of $\rho$ establishes namespace integrity, immutability, and the identity of the candidate objects under audit. It does not assert that the candidate output equals $\mathsf{Eval}(P,X)$, that its manifest is complete, or that a submitted semantic relation is true. The static workflow authorization binds the consumer file hash and frozen workflow scope but deliberately does not sign release-specific realization or payload fields; the keeper derives those fields from the unsigned proposer envelope and registered state. CertifiedLocal state jointly binds the canonical file hash, canonical stream hash, and KZG root. The current prototype still trusts registration/certification to establish file--vector and KZG--Merkle co-reference; it does not provide a cryptographic equality proof between independently supplied commitments.

The remaining parties are artifact producers, an authorized publisher, an
untrusted realization proposer, provers or certifiers, replicated keepers,
watchers, and a data-availability provider. The proposer may be fully malicious.
The optimistic model also permits a semantic-signing quorum to sign a false
local statement: BLS authenticates its members, exact digest, authorization,
and economic responsibility, but does not itself establish local semantic
truth. We therefore make watcher competence explicit. The premise
$\mathsf{WatcherComplete}_\rho(\ell)$ requires that, for every false
CertifiedLocal statement bound to release $\ell$, an honest watcher can obtain
the retained material, detect the falsehood, construct and submit a valid typed
challenge by the deadline, and have it included before finalization. Final
safety additionally assumes a sound and complete local fraud-proof judge and
the configured CometBFT safety and liveness threshold.

The adversary may alter backend choices, group membership, payload association,
message-carried RelationIR copies, workflow submissions, DA tuples, and replay
attempts. It may omit or overlap obligations, merge incompatible groups, reuse a
valid object across task/release/epoch boundaries, prove an unrelated canonical
statement, withhold challenge material, or submit malformed disputes. It cannot
rewrite an existing authority, break the assumed cryptographic primitives, or
exceed the consensus threshold. Bugs in semantic registration, supported
lowering, canonicalization, file/vector co-reference, requirement construction,
exact binding, and local interpretation are explicit TCB failures rather than
cryptographic error terms.

\subsection{Structural and Semantic Safety Events}
Let $R=\Req(\rho)$ and, when normalization is defined, let
$M=\Norm(\Gamma)$,
$G=\mathsf{Occ}(M)/{\sim_{\rm mat}}$, and
$\mathcal O_\rho(\mathcal E,\mathcal K)$ be the actual typed evidence-object set
reconstructed from submitted payloads and keeper state. Let
$\eta=(\Gamma,\mathcal E,\mathcal K)$. The registered inputs are the unique
canonical tensor stream $X=X(\rho)$ selected by the authenticated input
descriptors and root mapping in $\rho$; neither the proposer nor an evidence
object may replace $X$. Define
\begin{equation}
\begin{split}
\RealizationGap:={}&\neg\EntryComplete_\rho(R,M)\\
 &\vee\neg\MaterializationComplete_\rho(R,M,G)\\
 &\vee\neg\ObjectBind_\rho(G,\mathcal E,\mathcal K).
\end{split}
\label{eq:realization-gap}
\end{equation}
If $\Norm(\Gamma)$ is undefined, structural acceptance is false and
$\AdmissionSOVE_\rho(\eta)$ cannot occur; the displayed event is therefore evaluated
only on the well-typed normalization branch.
Here $\StructureAccept_\rho(\Gamma,\mathcal E,\mathcal K)$ means that the
requirement-derived entry, group, and actual-object phases of \CheckPlan{} accept.
Let $\chi=(\Gamma,\mathcal E,\mathcal K,A_{\mathrm{out}})$, where $\mathcal K$
is the authenticated release-scoped keeper-resolution snapshot. Let
$\mathsf{Recorded}_\rho(\ell,\chi,s)$ mean that release $\ell$'s immutable
AdmissionRecord in $s$ binds the authority-scoped manifest, realization,
materialization, evidence/certificate, workflow, and output commitments derived
from exactly $\chi$.
We separate the structural admission event from semantic release correctness:
\begin{equation}
\begin{aligned}
\AdmissionSOVE_\rho(\eta):={}&
\StructureAccept_\rho(\Gamma,\mathcal E,\mathcal K)\\[-0.25ex]
&\land\RealizationGap,\\
\FalseRelease_\rho(\ell,\chi,s):={}&
\Reach_\rho(s)\land\mathsf{Recorded}_\rho(\ell,\chi,s)\\
&\land\Consumable_\rho(A_{\mathrm{out}},\ell,s)\\
&\land\bigl(\neg\mathsf{AuditOK}_\rho(\Gamma,\mathcal E,A_{\mathrm{out}})\\[-0.25ex]
&\qquad\vee A_{\mathrm{out}}\neq\mathsf{Eval}(P,X)\bigr).
\end{aligned}
\label{eq:safety-events}
\end{equation}
The predicate $\AdmissionSOVE_\rho$ is a deterministic exact-realization
invariant relative to the
authenticated authority and structural checker TCB; it does not depend on
whether a bound evidence statement is true. The predicate $\FalseRelease_\rho$
is the conditional
semantic event and additionally depends on backend soundness, registered
co-reference, DA, watcher, inclusion, fraud-judge, and consensus assumptions.
Moving a false claim to \State{Provisional} is not a false release because that
state is non-consumable and remains challengeable.

\subsection{Non-Goals}
The core result excludes legal authorship, model quality, arbitrary
floating-point behavior, and unrestricted framework-to-RelationIR equivalence.
Optimistic local relations retain their stated data-availability, watcher,
judge, and inclusion assumptions.

\section{H-CRSPV Design}
\label{sec:design}

\subsection{System Overview}
\begin{figure*}[t]
  \centering
  \input{figures/hcrspv_pipeline_tikz.tex}
  \caption{H-CRSPV ownership and state flow. (a) The proposer selects a
  late-bound realization and proposes evidence sharing through complete keys;
  validators reconstruct the quotient, validate every group under registered
  policy, bind concrete objects, and then run native verification. (b) Only the
  guarded atomic finalizer changes a verified release into an active release;
  dashed branches show rejection, expiry, or an upheld challenge from
  \State{Pending} or \State{Provisional}.}
  \label{fig:pipeline}
\end{figure*}
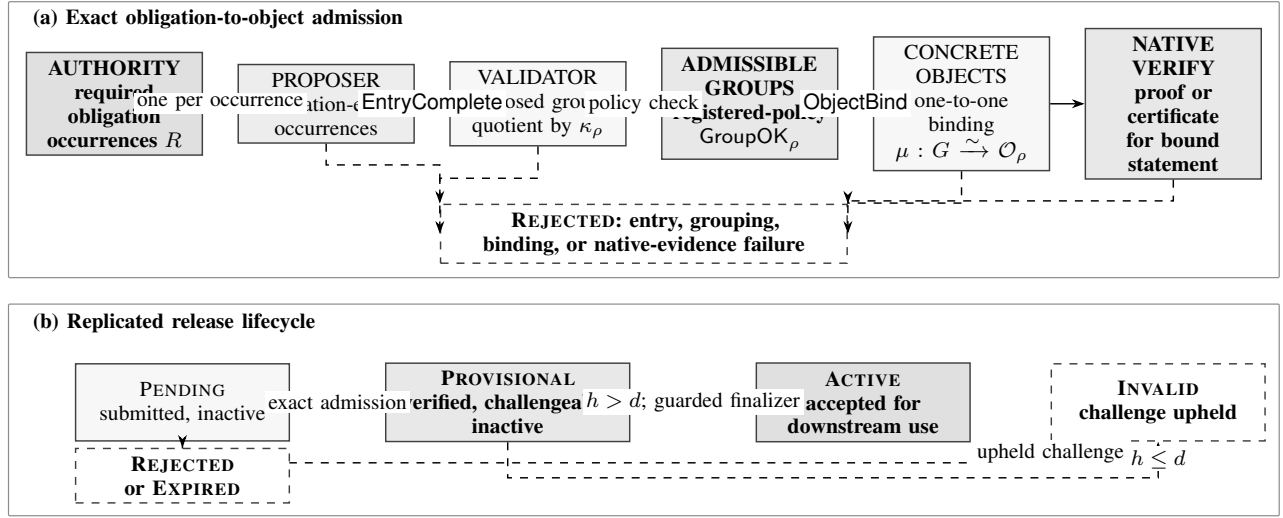

H-CRSPV enforces one invariant from registration to activation: each obligation
occurrence maps to one realization-entry occurrence; the proposer expresses
possibly singleton evidence-sharing groups through complete keys; validators
reconstruct and policy-check those groups; each admissible group binds one
concrete evidence object; and only a finalized release is active.
Figure~\ref{fig:pipeline} assigns each step to a single owner.

Task creation authenticates a versioned authority $\rho$ containing the
transformation profile and program, output identities, requirement root,
backend/SRS/DA policy, certifier set, challenge window $W$, and workflow key.
The requirement constructor derives $R=\Req(\rho)$ without reading the
proposal. An untrusted proposer later emits a finite realization multiset
$\Gamma$ using the evidence available at release time; complete keys in those
entries propose the concrete grouping. Every validator normalizes $\Gamma$,
reconstructs the resulting quotient, checks each proposed group against the
registered batching relation, resolves the release-scoped objects from payloads
and replicated keeper state, and only then invokes native evidence interfaces.
Backend verifiers establish facts about these resolved objects; they neither
select obligations, choose the grouping, nor activate a release.

We use the state names \State{Pending}, \State{Provisional}, and
\State{Active} throughout. \State{Active} is always paired with an accepted
AdmissionRecord; no consumer-facing query treats \State{Pending} or
\State{Provisional} as usable.

\subsection{Exact-Realization Contract}
The complete authenticated authority is
\begin{equation*}
\begin{aligned}
\rho=(&\mathit{ctx},\Pi_v,P,\Sigma,C,D,I,A_{\mathrm{out}},\\
      &\mathit{SRS},\mathit{DA},\mathit{CertSet},e,t_{\rm cert},W,\\
      &\mathit{Workflow}).
\end{aligned}
\end{equation*}
It covers the execution context, closed profile and typed DAG, policy and
catalogs, tensor and commitment descriptors, consumer output, production
setup, DA policy, certifier epoch and quorum, the authority-frozen window $W$,
and workflow layout and key. Given submitted realizations $\Gamma$ and evidence
$\mathcal E$, a component-only stack, on the defined normalization branch, computes
\begin{equation}
\CompAccept(\rho,\Gamma,\mathcal E)
 = \bigwedge_{r\in \Norm(\Gamma)} \mathsf{Verify}_{\rho}(r,\mathcal E_r),
\end{equation}
which validates received objects but does not establish equality with the
authority-derived requirement set.

We distinguish six objects throughout the paper. The \emph{requirement set}
$R$ is the authority-derived multiset. An \emph{obligation} $o\in R$ is one
required typed predicate. A \emph{realization entry} is one proposed evidence
assignment for one obligation occurrence. A \emph{materialization group}, or
\emph{shared-evidence group}, is a possibly singleton equivalence class of
occurrence-tagged entries with the same complete materialization key. It is
admissible only if the registered batching relation authorizes its exact
obligation multiset to share one evidence object. A \emph{concrete evidence
object} is the payload or certificate resolved from replicated state. Its
\emph{statement} is the canonical claim verified by the selected backend.

Normalization is a partial, fail-closed operation: if parsing, typing, or
canonicalization fails, structural acceptance is false. Otherwise it produces
the entry multiset $M=\Norm(\Gamma)$. Let
$\mathsf{Occ}(M)$ be its finite set of occurrence-tagged entries, so equal
entries remain distinct occurrences. Occurrence tags are assigned by the
canonical entry encoding and a duplicate ordinal after canonical sorting; they
are therefore deterministic and cannot be chosen by the proposer. Let
$\mathcal U_\rho$ be the finite typed universe of canonical obligation
identities generated by the registered DAG, block partition, profile, and
admissible relation families under $\rho$. The projection
$\pi^\rho_{\rm obl}:\mathsf{Occ}(M)\rightarrow\mathcal U_\rho$ removes
materialization fields and returns the obligation identity encoded by an entry.
For an occurrence subset $Q\subseteq\mathsf{Occ}(M)$,
$(\pi^\rho_{\rm obl})_\#Q$ denotes its pushforward multiset. Let
$\mathsf{EntryBind}_\rho(m)$ be the canonical semantic-binding tuple encoded
by $m$. We define $\mathsf{ExactEntryBind}_\rho(m,o)$ to require
$\mathsf{EntryBind}_\rho(m)=\mathsf{Bind}_\rho(o)$ and the entry's backend,
version, and assurance profile to belong to the authority's allowlist for $o$.
Exact entry coverage requires one correctly bound entry occurrence for every
obligation occurrence and no extras:
\begin{equation}
\begin{aligned}
&\EntryComplete_\rho(R,M):=\\[-0.25ex]
&\quad\bigl(\forall o\in\mathcal U_\rho:\;
\mathsf{mult}_{(\pi^\rho_{\rm obl})_\#\mathsf{Occ}(M)}(o)
 =\mathsf{mult}_R(o)\bigr)\\
&\quad\land\bigl(\forall m\in\mathsf{Occ}(M):
 \mathsf{ExactEntryBind}_\rho(m,\pi^\rho_{\rm obl}(m))\bigr).
\end{aligned}
\label{eq:entry-complete}
\end{equation}
This obligation--entry bijection rejects omission, duplicate coverage,
unmatched entries, assurance downgrade, and identity substitution without
requiring one proof object per obligation.

For an entry occurrence $m$, let $\kappa_\rho(m)$ be its complete
\textsf{MaterializationKey}: backend/version, actual handle, task, release,
candidate, artifact, output, epoch, profile, statement schema, and registered
batching class. Define
$m\sim_{\rm mat}m'\Longleftrightarrow\kappa_\rho(m)=\kappa_\rho(m')$.
Because it is induced by equality, $\sim_{\rm mat}$ is reflexive, symmetric,
and transitive. By choosing the allowed materialization-key fields in its
entries, the proposer proposes a concrete evidence-sharing partition. Each
validator deterministically reconstructs the occurrence-level quotient
$G=\mathsf{Occ}(M)/{\sim_{\rm mat}}$. At this point the classes in $G$ are
\emph{proposed groups}, not yet admissible groups. For each class $g\in G$,
define its exact obligation multiset as
$O_g=(\pi^\rho_{\rm obl})_\#g$; the common value of $\kappa_\rho$ on $g$ is
$\mathsf{MaterializationKey}_\rho(g)$. Then
\begin{equation}
\begin{aligned}
&\MaterializationComplete_\rho(R,M,G):=\\[-0.25ex]
&\quad G=\mathsf{Occ}(M)/{\sim_{\rm mat}}\\
&\quad\land\mathsf{OccPart}(\mathsf{Occ}(M),G)\\
&\quad\land\biguplus_{g\in G}O_g=R
 \land\bigwedge_{g\in G}\mathsf{GroupOK}_\rho(g).
\end{aligned}
\label{eq:materialization-complete}
\end{equation}
Here \textsf{OccPart} states that every entry occurrence belongs to exactly one
class. A proposed group becomes \emph{admissible} only when
$\mathsf{GroupOK}_\rho(g)$ holds. That predicate evaluates the complete
obligation multiset $O_g$ under the authority-registered batching relation and
authorizes every occurrence in $g$ to share the one materialization named by
their common key. It does not bind or verify the evidence object. Unknown,
mixed-assurance, or otherwise incompatible combinations evaluate to false.
Singleton classes are valid proposed groups and require the same policy check.
Thus batching can reduce object count only through a typed, complete,
occurrence-disjoint partition; repeated obligation identities retain their
multiplicity.

The resolver constructs $\mathcal O_\rho(\mathcal E,\mathcal K)$ from all
typed payloads supplied or referenced by the release and all authenticated
release-scoped certificates; any malformed, duplicate-key, or untyped object
causes resolution to fail rather than disappearing from this set. For each
group, its binding context is
\begin{equation}
\begin{aligned}
\mathsf{GroupBind}_\rho(g):={}&
 \bigl(\mathsf{Task},\mathsf{Release},\mathsf{Candidate},\\[-0.25ex]
&\mathsf{Artifact},\mathsf{Output},\mathsf{Epoch},\mathsf{Profile},\\[-0.25ex]
&\mathsf{Backend},\mathsf{SRS},\mathsf{DA},\\[-0.25ex]
&\mathsf{MaterializationKey}_\rho(g)\bigr).
\end{aligned}
\label{eq:group-bind}
\end{equation}
After every proposed group is admissible, the concrete-object contract requires
a bijection and two independent equalities:
\begin{equation}
\begin{aligned}
&\ObjectBind_\rho(G,\mathcal E,\mathcal K):=\\[-0.25ex]
&\quad\exists\,\mu:G\xrightarrow{\sim}\mathcal O_\rho(\mathcal E,\mathcal K)
  \;\text{s.t.}\\
&\quad\forall g\in G:\\[-0.25ex]
&\qquad \mathsf{Statement}(\mu(g))
 =\mathsf{Canon}_\rho(O_g,\mathsf{GroupBind}_\rho(g))\\[-0.25ex]
&\qquad\land\;
 \mathsf{ObjectKey}_\rho(\mu(g))\\[-0.25ex]
&\qquad\quad=\mathsf{MaterializationKey}_\rho(g).
\end{aligned}
\label{eq:object-bind}
\end{equation}
The first equality binds semantic content; the second binds the concrete object
that carries it. Totality assigns one object to every group, injectivity forbids
cross-group object reuse, and surjectivity forbids resolved orphan objects.
Together the checks reject missing payloads, proof swaps, unrelated statements,
stale handles, and cross-task, cross-release, cross-candidate, cross-epoch, or
cross-output reuse.

The evaluated profile illustrates the distinction. Its
\FinalCertifiedLocalEntries{} CertifiedLocal entry occurrences carry one
complete materialization key and therefore form one proposed quotient class.
\textsf{GroupOK} admits that class only if the registered policy permits its
exact $O_g$ to share a certificate. The bijection $\mu$ then binds the class to
one keeper-resolved BLS certificate whose relation-set digest commits to the
canonical statement for exactly the corresponding $O_g$. BLS verification
authenticates the 3-of-4 quorum statement; semantic truth still depends on the
registered data-availability, watcher, challenge, judge, and finalization
assumptions.

Let $\widehat\tau_{\rho,\Gamma,\mathcal E,A_{\mathrm{out}}}$ be the candidate
typed trace determined by authenticated registration and admitted evidence.
For $o\in R$, the judgment
$\widehat\tau_{\rho,\Gamma,\mathcal E,A_{\mathrm{out}}}\models o$ means that
the exact operational predicate bound by $o$ holds. We define
\begin{equation*}
\mathsf{AuditOK}_\rho(\Gamma,\mathcal E,A_{\mathrm{out}})
 :=\forall o\in R,\;
 \widehat\tau_{\rho,\Gamma,\mathcal E,A_{\mathrm{out}}}\models o.
\end{equation*}
\CheckPlan{} establishes that every obligation has a legal, object-bound
realization. \VerifyEvidence{} establishes predicate truth only under the
selected backend and its stated assumptions.

\subsection{Lifecycle and Consumability}
Candidate submission creates a \State{Pending} release under a unique
$(\mathit{task},\mathit{release})$ identity. Admission performs authority
lookup, requirement reconstruction, entry exactness, quotient reconstruction,
\textsf{GroupOK}, object binding, workflow binding, canonical-statement
derivation, and native verification in that order. Success creates only a
non-consumable \State{Provisional} record.

Certificate registration and evidence admission must each leave the
authority-defined window $W$ before the challenge deadline $d$:
\begin{equation*}
\max(h_{\rm cert},h_{\rm verify})+W\leq d.
\end{equation*}
A valid typed challenge through $d$ moves the release to \State{Invalid};
deterministically malformed evidence moves \State{Pending} to
\State{Rejected}; and incomplete certificate-bearing admission expires after
$d$. When the deadline has elapsed, every group is eligible, and no dispute
remains, the finalizer atomically writes \State{Active}, the accepted
AdmissionRecord, finalized certificate states, and bond settlement. Replay and
immutability guards prevent duplicate settlement or terminal-state mutation.
A replacement is a new release and cannot shorten the prior challenge window.

\subsection{Requirement Construction and RelationIR}
The closed \Pzero profile is a finite typed DAG. Its versioned profile is
\begin{equation*}
\begin{aligned}
\Pi_v=(&\mathit{OpSet},\mathit{TypeRules},\mathit{SemRules},\\
       &\mathit{ReqRules},\mathit{CanonRules},\mathit{EvidenceRules}).
\end{aligned}
\end{equation*}
The profile admits three node families:
\begin{equation*}
\begin{split}
n ::=~& \mathsf{AffineCombine}((a_i,d_i)_i)\\
 &\mid \mathsf{LocalMap}(g,d)\mid \mathsf{Reindex}(\pi,d),\\
g ::=~& \mathsf{DropRescaleMask}\mid\mathsf{ElementwiseScale}\\
 &\mid \mathsf{RoundDiv}\mid\mathsf{QuantizeUniform}.
\end{split}
\end{equation*}
Tensor descriptors bind shape, layout, fixed-point scale, signed width, range,
field, domain, and commitment layout. \textsf{RangeOK} is derived by the
checker, and \textsf{Reindex} covers chunk-aligned permutations. Dependencies
must point to earlier typed nodes, maps must be total, and every output must be
reachable. The production profile excludes unknown operators and
\texttt{DENSITY} sampling.

For example, the running adapter merge contains drop-rescale, affine, and
rounding nodes:
\begin{align*}
Z_i &= \mathsf{DropRescale}(\Delta_i;m_i,p_i),\\
N &= S_w X_0+a_A Z_A+a_B Z_B,\\
Y &= \mathsf{RoundDiv}(N,S_w).
\end{align*}
The constructor emits obligations for the affine relation, signed range,
rounding rule, registered reindexing, dependencies, and consumer output.
Writing $n_1,\ldots,n_m$ for topological order,
\begin{equation}
\begin{aligned}
N_{\rho}&=\biguplus_{j=1}^{m}\mathsf{NodeReq}(n_j,\rho),\\
\Req(\rho)&=\mathsf{CanonSort}(N_{\rho}\uplus
                                  \mathsf{OutputReq}(\rho)).
\end{aligned}
\label{eq:reqderive}
\end{equation}
Construction depends only on the registered authority, not on $\Gamma$.

Every entry is bound to a canonical semantic identity
\begin{equation}
\begin{split}
\mathsf{Bind}_\rho(o)=(&H(\rho),\mathit{nodeId},\mathit{rel},\mathit{prim},\\
 &(\mathit{inCommit},\mathit{inBlock})_{i},\\
 &\mathit{outCommit},\mathit{outBlock},H(\mathit{params}),\\
 &H(\mathit{deps}),\mathit{interpV}).
\end{split}
\label{eq:key}
\end{equation}
$H(\rho)$ covers the profile, program, descriptors, root mapping, policy, and
consumer-visible output. Block coordinates come from verified registered
descriptors rather than witness-selected fields. Consequently, a valid object
for one program, parameter set, dependency, coordinate range, commitment
domain, or interpreter cannot satisfy another obligation.

\subsection{Restricted Frontend and Native Backends}
For the evaluated declarative PEFT/Safetensors LoRA subset, a restricted
frontend compiles source metadata and tensor headers into a typed program and a
lowering certificate:
\begin{equation}
\begin{aligned}
\mathsf{Compile}(S)&\rightarrow(P,\pi_{\mathrm{lower}}),\\
\mathsf{CheckLower}(H(S),P,\pi_{\mathrm{lower}})&=1.
\end{aligned}
\end{equation}
The certificate binds the base revision, immutable tensor hashes, rank, alpha,
orientation, target modules, shapes, dtypes, library versions, consumer port,
and output map. \textsf{CheckLower} reparses these fields before realization
and compares the \Pzero reference output with the registered declarative source
semantics.

The evidence layer has mandatory deterministic bindings, production KZG for the
global affine statement, and direct or BLS-certified local relations. The
keeper derives every statement from authority-owned RelationIR and accepts only
registered backend, SRS, and DA profiles. A CertifiedLocal aggregate proves that
three distinct authorized members signed the exact group statement; semantic
truth additionally relies on retained challenge data, watcher completeness, and
the typed fraud-proof judge. The mixed KZG/Merkle path binds both roots through
one authenticated registrar transcript. This registrar supplies the explicit
co-reference premise used by Proposition~1.

The checker performs Eqs.~\eqref{eq:entry-complete}--
\eqref{eq:object-bind} before expensive backend dispatch. Unknown backend
labels, malformed transcripts, policy downgrades, unrelated statements, and
DA/SRS mismatch reject. Here $\mathsf{EvidenceOK}_\rho$ means that each
keeper-resolved object is accepted by its registered native interface, while
$\mathsf{WorkflowOK}_\rho$ binds the candidate and consumer output to the
registered release workflow.

For comparison, on the defined normalization branch, let
$\chi=(\Gamma,\mathcal E,\mathcal K,A_{\rm out})$,
$R=\Req(\rho)$, $M=\Norm(\Gamma)$, and
$G=\mathsf{Occ}(M)/{\sim_{\rm mat}}$. The fixed inspector and the
keeper acceptance predicates at a committed consensus state $s$ are
\begin{equation*}
\begin{aligned}
\mathsf{InspectorOK}_\rho(\Gamma)
  &:={\mathsf{ReqIds}}(\Req(\rho))
     ={\mathsf{ReqIds}}(\Norm(\Gamma)),\\
\mathsf{HAccept}_\rho(\ell,\chi,s)
  &:=\Reach_\rho(s)\\
  &\quad\land\mathsf{Recorded}_\rho(\ell,\chi,s)\\
  &\quad\land\EntryComplete_\rho(R,M)\\
  &\quad\land\MaterializationComplete_\rho(R,M,G)\\
  &\quad\land\ObjectBind_\rho(G,\mathcal E,\mathcal K)\\
  &\quad\land\mathsf{WorkflowOK}_\rho(\ell,\Gamma,A_{\rm out},s)\\
  &\quad\land\mathsf{EvidenceOK}_\rho(\ell,G,\mathcal E,\mathcal K,s)\\
  &\quad\land\FinalAccept_\rho(\ell,s).
\end{aligned}
\end{equation*}
$\mathsf{ReqIds}$ returns a multiplicity-preserving identifier multiset; the
$\mathsf{Recorded}_\rho$ predicate above prevents substituting a different
manifest, object set, keeper state, or output after admission.
$\mathsf{InspectorOK}$ compares identifiers; it neither proves field-level
entry binding nor owns grouping, object resolution, or consumability.
$\mathsf{HAccept}$ composes those checks with the final-acceptance predicate
formalized in Sec.~\ref{sec:security}.

\section{Security Analysis}
\label{sec:security}

The security argument has three layers (Table~\ref{tab:guarantee-ladder}).
Each layer adds assumptions rather than silently promoting a structural set
check into an end-to-end semantic claim. The closed \Pzero specification
provides the grammar, type and obligation rules, canonical identities, and
backend admissibility policy used below.

\begin{table}[t]
\centering
\caption{Guarantee ladder: conclusions and additional assumptions.}
\label{tab:guarantee-ladder}
\TableFont
\begin{tabularx}{\columnwidth}{@{}L{0.19\columnwidth}L{0.32\columnwidth}Y@{}}
\toprule
Layer & Conclusion & Additional assumptions \\
\midrule
Structural & No omission, overlap, orphan, or cross-binding & Authority; constructor; normalizer; grouper; resolver \\
Finalized semantic & Consumable output satisfies the registered program & Structural layer; backend adequacy; co-reference; DA; watcher; judge; consensus \\
Source-level TCB transfer & Output satisfies the supported source semantics & Semantic layer; trusted checked-frontend correspondence \\
\bottomrule
\end{tabularx}
\end{table}

\subsection{Coverage, Binding, and Finalization}
Fix an authority version $\rho$. Its replicated lifecycle is the labeled
transition system
\begin{equation*}
\mathcal T_\rho=(\mathcal S_\rho,s^0_\rho,\Delta_\rho),\qquad
\Delta_\rho\subseteq\mathcal S_\rho\times\mathcal A_\rho
\times\mathcal S_\rho.
\end{equation*}
We write $u\xrightarrow{a}_\rho v$ for $(u,a,v)\in\Delta_\rho$ and
$\Reach_\rho(s)$ iff there exist $k\geq0$, states
$s_0=s^0_\rho,\ldots,s_k=s$, and actions $a_0,\ldots,a_{k-1}$ with
$s_i\xrightarrow{a_i}_\rho s_{i+1}$ for every $i<k$. We write
$u\preceq_\rho s$ when a (possibly empty) $\Delta_\rho$ path leads from $u$ to
$s$. Each accepted transaction is one action in the CometBFT total order,
including transactions within the same block.

The initial state contains no active release or accepted AdmissionRecord.
Only admission may create \State{Provisional}; only the finalizer may create
\State{Active} with an accepted record. No transition admits a challenge from
terminal \State{Active} or reopens its dispute counter.

The lifecycle predicates deliberately separate exact admission, eligibility to
finalize, and final acceptance. Let $\mathsf{ExactAdmission}_\rho(\ell,s)$
abbreviate the conjunction of Eqs.~\eqref{eq:entry-complete}--\eqref{eq:object-bind}
with the explicit \textsf{WorkflowOK} and \textsf{EvidenceOK} arguments under
release $\ell$'s registered context. Admission stores the canonical inputs,
derived roots, group set, object keys, check results, and output identity in one
immutable AdmissionRecord; later states read this record rather than
re-evaluating mutable external data. Let $G_\ell$ be the immutable admitted
group set. Then
\begin{equation}
\begin{aligned}
\FinalizeGuard_\rho(\ell,s)&:={}\\[-0.5ex]
 &\mathsf{State}_s(\ell)=\texttt{PROVISIONAL}\\
 &\land\mathsf{ExactAdmission}_\rho(\ell,s)\\
 &\land\mathsf{WindowBound}_\rho(W,\ell,s)\\
 &\land h(s)>d_\ell\\
 &\land\mathsf{AllEligible}_\rho(G_\ell,\ell,s)\\
 &\land\mathsf{NoDispute}_\rho(\ell,s),\\[-0.25ex]
\mathsf{NoDispute}_\rho(\ell,s)&\Longleftrightarrow
 \mathsf{OpenDisputeCount}_s(\ell)=0.
\end{aligned}
\label{eq:finalize-guard}
\end{equation}
Every predicate is evaluated under the same registered task, release,
candidate, artifact, output, epoch, profile, backend, SRS, DA, and challenge
context. Here \textsf{WindowBound} requires both certificate registration and
evidence verification to leave the authority-frozen $W$ blocks before the
deadline. A valid challenge action is enabled while $h(s)\leq d_\ell$ and moves
\State{Pending} or \State{Provisional} to terminal \State{Invalid}; a finalize
action is enabled only when $h(s)>d_\ell$. Thus a challenge included at
$h=d_\ell$ is ordered and applied before any later enabled finalizer. When
Eq.~\eqref{eq:finalize-guard} holds,
$s\xrightarrow{\mathsf{finalize}(\ell)}_\rho s'$ atomically establishes
\begin{equation*}
\begin{aligned}
\mathsf{State}_{s'}(\ell)&=\texttt{ACTIVE},\\
\mathsf{AdmissionRecord}_{s'}(\ell)&=\texttt{ACCEPTED},\\
&\text{for every }c\in\mathsf{Certs}(\ell):\\[-0.5ex]
&\qquad\mathsf{CertState}_{s'}(c)=\texttt{FINALIZED},\\[-0.5ex]
&\qquad\mathsf{BondState}_{s'}(c)=\texttt{AVAILABLE}.
\end{aligned}
\end{equation*}
No other action may establish either active or accepted, and terminal release,
certificate, and settlement fields are immutable. Define the historical
finalization predicate and current final acceptance as follows; for compactness,
$\mathsf{AdmRec}_s(\ell)$ abbreviates
$\mathsf{AdmissionRecord}_s(\ell)$.
\begin{equation}
\begin{aligned}
\mathsf{Finalized}_\rho(\ell,s):={}&\exists u,v:\;
 u\xrightarrow{\mathsf{finalize}(\ell)}_\rho v\\[-0.25ex]
&\land\Reach_\rho(u)\land v\preceq_\rho s\\[-0.25ex]
&\land\mathsf{AdmRec}_v(\ell)=\texttt{ACCEPTED},\\
\FinalAccept_\rho(\ell,s):={}&
 \mathsf{State}_s(\ell)=\texttt{ACTIVE}\\[-0.25ex]
&\land\mathsf{AdmRec}_s(\ell)=\texttt{ACCEPTED},\\
\Consumable_\rho(A_{\mathrm{out}},\ell,s):={}&
 \mathsf{ReleaseOutput}_s(\ell)=A_{\mathrm{out}}\\[-0.25ex]
&\land\FinalAccept_\rho(\ell,s).
\end{aligned}
\label{eq:final-accept}
\end{equation}
Only the atomic finalizer may establish Eq.~\eqref{eq:final-accept}, and it may
do so only when the guard in Eq.~\eqref{eq:finalize-guard} holds. Thus
\State{Provisional} is never consumable.

\textbf{Determinism Lemma (Replica-consistent reconstruction).} For a fixed
authority snapshot $\rho$, canonical proposal bytes, evidence bytes, and
release-scoped keeper snapshot, every correct replica computes the same
$R$, $M$, $G$, resolved object set, and structural decision.

\emph{Proof sketch.} Requirement construction and normalization are total
deterministic functions on their accepting domains; all sets are ordered by
canonical encodings; occurrence tags use canonical order and duplicate
ordinals; equality of the complete materialization key induces a unique
quotient; and the resolver reads the same authenticated state snapshot. No
consensus decision depends on host map iteration or arrival order. The complete
argument appears in Appendix~\ref{app:proofs}. \hfill$\square$

\textbf{Lemma 1 (Exact Obligation Coverage).} For authenticated $\rho$, let
$R=\Req(\rho)$ and $M=\Norm(\Gamma)$. If the entry phase of \CheckPlan{} accepts,
then $\EntryComplete_\rho(R,M)$ holds.

\emph{Proof sketch.} The checker derives $R$ independently of $\Gamma$ and
enforces multiplicity equality plus exact field binding for every entry.
Appendix~\ref{app:proofs} gives the accepting-path argument. \hfill$\square$

\textbf{Lemma 2 (Materialization Binding).} If the materialization phase
accepts after Lemma~1, then materialization completeness
(Eq.~\eqref{eq:materialization-complete}) and object binding
(Eq.~\eqref{eq:object-bind}) both hold.

\emph{Proof sketch.} Complete keys in the entries induce the proposer's unique
proposed quotient. Each validator reconstructs that quotient, checks its
disjoint complete union, applies the registered batching relation, and resolves
exactly one context- and statement-bound object per admissible group. Detailed
branch coverage appears in
Appendix~\ref{app:proofs}. \hfill$\square$

\textbf{Bridge Lemma (Operational--abstract correspondence).}
\label{lem:bridge}
Let
$\OpRep_{\rho,\ell}(q,\mathcal K)=(\Gamma,\mathcal E,A_{\mathrm{out}})$
mean that $q$ is the release submission for $(\rho,\ell)$, contains
realization proposal $\Gamma$, binds candidate output $A_{\mathrm{out}}$, and
selects the supplied or authenticated-reference inputs $\mathcal E$ when
evaluated against authenticated keeper snapshot $\mathcal K$. If
$\EntryComplete$, $\MaterializationComplete$, and $\ObjectBind$ hold for this
same operational instance, the unique bijection $\mu$ in
Eq.~\eqref{eq:object-bind} instantiates
\begin{equation}
\begin{aligned}
\SubObj(q)&:=\mathcal O_\rho(\mathcal E,\mathcal K),\\
\Cov_{\rho,\ell}(q,\mu(g))&:=O_g \qquad(g\in G),
\end{aligned}
\label{eq:operational-cov}
\end{equation}
and
\begin{equation}
\begin{split}
&\OpRep_{\rho,\ell}(q,\mathcal K)
  =(\Gamma,\mathcal E,A_{\mathrm{out}})\\[-0.25ex]
&\quad\land\EntryComplete\land\MaterializationComplete\\[-0.25ex]
&\quad\land\ObjectBind
\Longrightarrow\CompleteRealization_{\rho,\ell}(q).
\end{split}
\label{eq:bridge}
\end{equation}

\emph{Proof sketch.} Surjectivity of $\mu$ defines coverage for every resolved
submitted object; injectivity and unique complete object keys make coverage
single-valued. Exact entry binding, $\mathsf{GroupOK}$, and the statement,
context, and object-key equalities make the coverage authority-scoped.
Reindexing the disjoint multiset union in
Eq.~\eqref{eq:materialization-complete} through $\mu$ gives
$\biguplus_{e\in\SubObj(q)}\Cov_{\rho,\ell}(q,e)=R_\rho$, proving both clauses
of Eq.~\eqref{eq:complete-realization}. Appendix~\ref{app:proofs} expands the
resolver branches. \hfill$\square$

\textbf{\ExactRealizationTheorem{} (Exact-realization invariant).}\label{thm:exact-realization}
Let
$\StructOK_\rho(\Gamma,\mathcal E,\mathcal K):=
\EntryComplete_\rho(R,M)\land
\MaterializationComplete_\rho(R,M,G)\land
\ObjectBind_\rho(G,\mathcal E,\mathcal K)$.
Relative to an authenticated authority and faithful execution of the registered
requirement constructor, normalizer, quotient reconstruction,
\textsf{GroupOK} logic, canonical encoder, and typed object resolver,
\begin{equation}
\StructureAccept_\rho(\Gamma,\mathcal E,\mathcal K)
\Longrightarrow\StructOK_\rho(\Gamma,\mathcal E,\mathcal K).
\label{eq:exact-realization}
\end{equation}
By definition, $\StructOK_\rho\Longleftrightarrow\neg\RealizationGap$ on the
well-typed normalization branch.
Consequently, for the same operational instance,
\begin{equation}
\begin{split}
&\OpRep_{\rho,\ell}(q,\mathcal K)
  =(\Gamma,\mathcal E,A_{\mathrm{out}})\\[-0.25ex]
&\quad\land\StructureAccept_\rho(\Gamma,\mathcal E,\mathcal K)\\[-0.25ex]
&\qquad\Longrightarrow\CompleteRealization_{\rho,\ell}(q),
\end{split}
\label{eq:structure-to-abstract}
\end{equation}
so both $\AdmissionSOVE_\rho(\eta)$ and the abstract
$\SOVEAccept^I_{\rho,\ell}(q)$ event are false for H-CRSPV under the structural
assumptions in Table~\ref{tab:guarantee-ladder}.

\emph{Proof sketch.} Lemma~1 establishes the obligation--entry bijection;
Lemma~2 establishes registered-policy acceptance of the proposal-induced
partition and the concrete-object bijection over the same authority snapshot.
Their conjunction is exactly $\StructOK_\rho$, equivalently the negation of
Eq.~\eqref{eq:realization-gap}; the Bridge Lemma then yields
Eq.~\eqref{eq:structure-to-abstract}. \hfill$\square$

\textbf{Lemma 3 (Reachable-state finalization safety).} For every release
$\ell$, state $s$, and registered output $A_{\mathrm{out}}$,
\begin{equation}
\begin{aligned}
&\Reach_\rho(s)\\[-0.25ex]
&\quad\land\Consumable_\rho(A_{\mathrm{out}},\ell,s)\\[-0.25ex]
&\quad\Longrightarrow\mathsf{Finalized}_\rho(\ell,s)\\[-0.25ex]
&\qquad\land\mathsf{ExactAdmission}_\rho(\ell,s)\\[-0.25ex]
&\qquad\land\mathsf{NoDispute}_\rho(\ell,s).
\end{aligned}
\label{eq:reachable-finalization}
\end{equation}
The witnessing finalizer ran after $d_\ell$, found every group eligible, and
atomically finalized certificates and settled each bond exactly once.

\emph{Proof sketch.} Induct on the path witnessing $\Reach_\rho(s)$. The initial
state is non-consumable. Admission, rejection, challenge, expiry, and replay
transitions cannot establish active-plus-accepted. The only inductive case that
can establish it is an enabled atomic finalizer; its guard supplies every
conjunct, and terminal immutability preserves them. The full transition
argument appears in Appendix~\ref{app:proofs}. \hfill$\square$

\subsection{Conditional Composition Guarantee}
The structural and lifecycle lemmas do not independently prove external source semantics or
backend soundness. For each registered primitive, we assume the per-node
adequacy condition. Let $V_j$ be the admitted value at node $n_j$ and $V_j^*$
its reference value in topological order. Writing $\mathsf{Pred}(j)$ for its
registered predecessors, adequacy is the typed implication
\begin{equation*}
\begin{aligned}
&\bigl(\forall i\in\mathsf{Pred}(j):V_i=V_i^*\bigr)
 \land\bigl(\forall o\in\mathsf{NodeReq}(n_j,\rho):
 \widehat\tau\models o\bigr)\\[-0.25ex]
&\qquad\Longrightarrow
 V_j=\mathsf{Sem}_\rho(n_j,(V_i^*)_{i\in\mathsf{Pred}(j)}).
\end{aligned}
\end{equation*}
Backend lifting is deliberately typed:
\begin{equation}
\begin{aligned}
\mathsf{Verify}^{\rm direct/KZG}_\rho(\mu(g))=1
 &\Longrightarrow \forall o\in O_g:
 \widehat\tau_{\rho,\Gamma,\mathcal E,A_{\mathrm{out}}}\models o,\\
\mathsf{Verify}^{\rm CL\mbox{-}BLS}_\rho(\mu(g))=1
 &\Longrightarrow
 \mathsf{QuorumSigned}_\rho(\mathsf{Stmt}(g)).
\end{aligned}
\label{eq:lifting}
\end{equation}
The second line authenticates an exact statement; it does not alone establish
that statement's semantic truth. For CertifiedLocal groups, truth enters the
composition argument only through finalization together with DA,
$\mathsf{WatcherComplete}_\rho(\ell)$, and fraud-judge assumptions.
The registered semantic statement for such a group is the conjunction of its
exact obligations, $\mathsf{Stmt}(g)\equiv\bigwedge_{o\in O_g}o$, encoded with
the binding context in Eq.~\eqref{eq:group-bind}.

\textbf{\SemanticCompositionProposition{} (Finalized semantic composition).}\label{prop:composition}
Let $P$ be a well-typed closed \Pzero\ DAG with registered inputs $X$, let
$\chi=(\Gamma,\mathcal E,\mathcal K,A_{\mathrm{out}})$, and suppose the immutable
AdmissionRecord for $\ell$ binds exactly $\chi$. Assume authentic and adequate task authority;
canonical decoding and registered file/vector co-reference; \ExactRealizationTheorem{} and
Lemma~3; backend soundness; a valid registered workflow;
per-node adequacy; challenge data available through the deadline;
$\mathsf{WatcherComplete}_\rho(\ell)$; a sound and complete fraud-proof judge; and
the configured CometBFT threshold. If $\Reach_\rho(s)$,
$\mathsf{Recorded}_\rho(\ell,\chi,s)$, and
$\Consumable_\rho(A_{\mathrm{out}},\ell,s)$ hold, then
\begin{equation}
\begin{aligned}
\Reach_\rho(s)\land{}&
 \mathsf{Recorded}_\rho(\ell,\chi,s)\\[-0.25ex]
{}\land{}&
 \Consumable_\rho(A_{\mathrm{out}},\ell,s)\\[-0.25ex]
\Longrightarrow{}&
 \mathsf{AuditOK}_\rho(\Gamma,\mathcal E,A_{\mathrm{out}})\\
 &\land A_{\mathrm{out}}=\mathsf{Eval}(P,X).
\end{aligned}
\label{eq:semantic-sufficiency}
\end{equation}

\emph{Proof sketch.} The record premise fixes the same $\chi$ throughout the
argument, and Lemma~3 supplies exact admission for that record. Lemma~1 supplies
every exact obligation in its manifest; Lemma~2 validates the proposed groups
and binds each admissible group to exactly one keeper-resolved object.
Equation~\eqref{eq:lifting} lifts direct/KZG object verification to obligation
truth and BLS verification only to an authenticated quorum statement.
Per-node adequacy and topological induction establish every intermediate and
output block. Output binding identifies the result with $A_{\mathrm{out}}$.
For an optimistic group, suppose its conjunction were false. Watcher
completeness supplies a timely included typed challenge, judge completeness
upholds it, and $\Delta_\rho$ moves the release to terminal \State{Invalid},
contradicting Lemma~3 and consumability. This supplies the remaining semantic
step. The result is a conditional
pen-and-paper composition argument, not mechanized verification of its TCB.
\hfill$\square$

Define $\mathsf{FrontendRel}_\rho(S,P,\pi_{\rm lower})$ as the trusted
correspondence invariant that the supported source parser, tensor and parameter
mapping, consumer port, and output map are field-complete and that the supported
source evaluator is observationally equivalent to $\mathsf{Eval}(P,\cdot)$ at
the bound output. \textsf{CheckLower} checks the registered hashes and fields;
it does not prove this invariant for arbitrary executable code.

\textbf{\SourceTransferClaim{} (Scoped frontend-TCB transfer).} Let $S$ belong
to the supported declarative PEFT/Safetensors subset. If \textsf{CheckLower}
accepts $(H(S),P,\pi_{\rm lower})$, $\mathsf{FrontendRel}_\rho$ holds, and the
premises of \SemanticCompositionProposition{} hold, then the accepted output
equals the registered source semantics. This is a transfer through an explicit
frontend TCB premise, not an independently proved source-level theorem, and it
does not extend to arbitrary Python execution.

\begin{table*}[!t]
\centering
\caption{Normative release transitions. Guards are evaluated from replicated
state; the listed release, certificate, and bond effects commit atomically.}
\label{tab:state}
\TableFont
\begin{tabularx}{\textwidth}{@{}L{0.20\textwidth}Y L{0.31\textwidth}C{0.09\textwidth}@{}}
\toprule
Transition & Guard or rejected action & Atomic release/certificate/bond effect & Consumable \\
\midrule
\State{Pending}$\rightarrow$\State{Provisional} &
$\mathsf{ExactAdmission}_\rho(\ell,s)\land h\leq d_\ell$ &
Store immutable exact-admission record; certificates remain provisional and bonds locked & No \\
\State{Pending}$\rightarrow$\State{Rejected} &
Complete deterministic evidence establishes a registered invalid state &
Write rejected record; release and configured settlement become terminal & No \\
\shortstack[l]{\State{Pending}/\State{Provisional}\\$\rightarrow$\State{Invalid}} &
Valid typed fraud proof and $h\leq d_\ell$ &
Revoke affected certificates and slash configured bonds exactly once & No \\
$s\rightarrow s$ &
Malformed/invalid typed challenge or unauthorized generic dispute &
Reject transaction; no release, certificate, or bond mutation & Unchanged \\
\shortstack[l]{Cert.-bearing \State{Pending}\\$\rightarrow$\State{Expired}} &
$h>d_\ell$ with no complete admission or open dispute &
Mark certificate \State{ExpiredUnused}; make bond available & No \\
\State{Provisional}$\rightarrow$\State{Active} &
$\FinalizeGuard_\rho(\ell,s)$ &
Atomically write accepted record, finalize all certificates, and release bonds once & Yes \\
Terminal $\rightarrow$ same terminal &
Replay or illegal terminal mutation &
Reject or return the recorded result; never repeat settlement & Unchanged \\
\bottomrule
\end{tabularx}
\end{table*}

\FloatBarrier

\subsection{SOVE Rejection}
\textbf{Corollary 1 (No admitted structural omission).} Under the structural
premises of \ExactRealizationTheorem{} and the $\OpRep$ relation for the same
admission instance, \StructureAccept{} implies
\CompleteRealization{} by Eq.~\eqref{eq:structure-to-abstract}. Any missing,
duplicated, overlapping, cross-bound, non-admissibly grouped, reused, or orphaned
obligation/object therefore prevents structural acceptance regardless of the
validity of submitted objects. Separately, Lemma~3 prevents a provisional,
disputed, or otherwise unfinalized release from satisfying
Eq.~\eqref{eq:final-accept}. Neither conclusion requires the semantic premises
of \SemanticCompositionProposition{}.

\subsection{Submitted-Object Interface Separation}
\textbf{Observation 1 (Submitted-object non-implication).} Component acceptance
over only submitted objects does not imply manifest completeness unless the
verifier also has a trusted complete requirement set, an equivalent
reconstruction mechanism, or a proof of statement-set equality.

\emph{Proof sketch.} Remove a critical obligation $o^*$ and provide valid
evidence for every remaining entry. The submitted-object view observes only
valid objects; a verifier that reconstructs the complete requirement set
rejects the missing $o^*$.

\subsection{Residual Risk}
Acceptance remains conditional on distinct trust domains: task-authority adequacy and immutability, source lowering for the supported subset, canonicalization, file/vector co-reference, realization-to-evidence association, cryptographic soundness, the consensus threshold, data availability, watcher completeness, and the local-interpreter TCB. We do not assign a numerical union bound because deterministic implementation bugs and deployment-liveness failures lack a common probability experiment in the present model.

\FloatBarrier

\section{Prototype and Evaluation}
\label{sec:implementation}
\label{sec:evaluation}

The implementation has one versioned admission schema with several evidence realizations. The off-chain path consists of the \Pzero profile, a restricted PEFT 0.18.1/Safetensors 0.8.0 frontend~\cite{huggingface2026peft0181,huggingface2026safetensors080}, an exact-binding manifest checker, a registrar-dependent co-reference gate, a Rust/Python heterogeneous dispatcher, commitment-bound local evidence, and consumer-output checks. The Cosmos SDK v0.50.5 keeper~\cite{cosmossdk2024v0505}, replicated by CometBFT v0.38.6~\cite{buchman2016tendermint,cometbft2024v0386}, loads an immutable task Verification Authority, verifies typed production-KZG and task-owned BLS evidence, checks an unsigned release envelope against a reusable task-registered static workflow authorization, and enforces the release lifecycle.

\subsection{Prototype and Experimental Setup}
The checker computes requirements before reading evidence verdicts. It rejects
malformed or ambiguous identities, multiplicity mismatch, missing or unexpected
entries, incorrect dependencies, stale policy versions, backend downgrades,
incomplete output binding, policy-invalid proposer grouping, and concrete-object
mismatch. The Go keeper loads authority from consensus state, derives and sorts
obligations, reconstructs the quotient induced by the entries' complete typed
materialization keys, evaluates \textsf{GroupOK}, and reconstructs objects from
mandatory/KZG/local payloads and CertifiedLocal state. A hash string alone can
neither create a payload nor authorize batching.

For certificate-bearing terminal paths, an active release has an accepted
record, a finalized certificate, and an available bond. An invalid release has
a revoked certificate and slashed bond; an expired release marks its unused
certificate expired and returns the bond. A malformed fraud proof changes none
of these states.

Only an uncertified and unbonded pending reservation may be cancelled,
replaced, or expired. The consumer API calls one predicate equivalent to
Eq.~\eqref{eq:final-accept}; queries do not expose provisional artifacts as
accepted. The legacy non-authority generic-dispute path permits at most one open
dispute and only the frozen governance authority may resolve it; terminal state
is immutable. Transaction handlers emit deterministic protocol facts only. Host
wall time, CPU time, and memory are recorded by the experiment harness rather
than consensus events.

The trusted path contains the parser, type checker, requirement constructor,
canonicalizer, object resolver and interpreters; lowering, authority, and the
current co-reference registrar are also trusted. Any mechanism deciding what
the constructor must require therefore remains in the TCB. \system bounds it with
a closed versioned language, hash-pinned authority,
fail-closed operators, replicated execution, and differential oracles; it does
not choose late realizations or require an online semantic planner. All planning
and evidence-production roles remain untrusted. Cryptography, BFT, DA, and
watcher liveness are separate assumptions; testing reduces but cannot remove
the TCB. After this interface is fixed, late actors cannot change obligations,
objects, or consumability.

\subsubsection{KZG and registered setup}
The bundled Rust KZG worker~\cite{kate2010kzg} implements chunked vector
commitments, openings, batch verification, and global affine checks. The keeper
permits only the registered production backend and SRS profile, both loaded from
consensus state. Transaction execution does not read node-local files or
environment variables. The affine statement is reconstructed from
authority-owned RelationIR; a valid proof for another relation therefore fails
the canonical-statement gate. The co-reference registrar remains a TCB and is
not presented as a KZG--Merkle equality proof.

\subsubsection{Certified local relations}
The BLS quorum~\cite{boneh2001bls} signs a canonical digest over domain/schema,
chain/task/release/candidate, publisher/submitter, artifact and output digests
and references, materialization-group relation set, DA tuple, certifier
set/epoch/bitmap, deadline, minimum remaining window $W$, and bond records. The keeper checks three distinct
members of four, bitmap consistency, epoch, PoP state, and aggregate pairing.
This authenticates responsibility for the exact local statement; semantic
truth remains optimistic. The frozen certificate's statement covers exactly
\FinalCertifiedLocalEntries{} local obligations and remains bond-locked through provisional/disputed
state. A valid fraud proof rejects and slashes; an unchallenged exact
certificate may be finalized after the deadline. The final negative report
separates malformed encoding, bad PoP, stale epoch, wrong bitmap, and a valid
aggregate over an alternate legal digest (\FinalWrongDigestResult).

\subsubsection{Registered data availability}
The authority freezes a versioned DA profile. The current loopback profile is
an implementation control, not external DA: it requires
$\mathsf{DACommitment}=\mathsf{ArtifactDigest}$ and
$\mathsf{DALocation}=\mathsf{CanonicalCIDOrArtifactRef}$. Candidate,
certificate, static workflow authorization, release envelope, provisional
record, final record, and BLS
digest bind the same tuple. Unknown versions, short retention, commitment or
location swaps, and cross-release tuple reuse fail closed.

\subsubsection{Cheap-check ordering}
Both handlers are structured to reject inexpensive malformed cases before cryptographic or interpreter work:
\begin{align*}
\text{size/capability} &\rightarrow \text{manifest binding} \\
&\rightarrow \text{registry/domain} \rightarrow \text{decode/subgroup} \\
&\rightarrow \text{crypto} \rightarrow \text{local judge} \rightarrow \text{state write}.
\end{align*}
This ordering matters because the checker itself becomes an attack surface once adversaries can submit manifests and challenges. The frozen campaign reports observable lifecycle and query results; rejection-path backend-call and state-write counters remain null and unobservable. Sustained throughput, external-network finality, and external-DA retention remain outside this paper's claim.

We evaluate six questions. RQ1 tests the obligation--entry--group--object
contract. RQ2 tests lifecycle safety. RQ3 asks whether one authority can support
post-registration evidence changes without a trusted online planner. RQ4 tests
source binding and compares fair completeness baselines. RQ5 follows one
artifact through replicated admission. RQ6 reports cost and deployment
boundaries. Appendix~\ref{app:validation-campaigns} contains the supporting
mutation, cross-language, trace, and local-judge campaigns.

\subsection{RQ1: Entry--Group--Object Exactness}
The fixed PEFT instance contains \FinalRealizationEntries{} required obligation
occurrences, and its honest proposal contains the same number of realization
entry occurrences. Those entries induce \FinalMaterializationGroups{}
groups---one \FinalCertifiedLocalEntries{}-entry group and 25 singleton
groups---each of which satisfies \textsf{GroupOK}; \ObjectBind{} then maps the
admissible groups one-to-one to \FinalActualEvidenceObjects{} concrete objects.
The large group shares one CertifiedLocal 3-of-4 aggregate certificate, so its
entries are neither independent proofs nor independent objects. This is the
frozen chain
\emph{required occurrences $\rightarrow$ proposal entries $\rightarrow$
proposal-induced groups $\rightarrow$ GroupOK-admissible groups
$\rightarrow$ keeper-resolved objects}.
Directed controls distinguish three failure owners: incompatible sharing fails
\textsf{GroupOK}; missing, reused, orphan, stale, or cross-bound objects fail
\textsf{ObjectBind}; and invalid proofs or certificates fail native
verification. Entry omissions, duplicates, and unmatched identities fail the
earlier exact-entry gate. Every legal and adversarial row produces the expected
decision; Appendix~\ref{app:matrix} preserves the campaign counts and
command-level evidence.

We additionally construct six incorrect endpoints with fresh, valid evidence
for every relation that remains declared. The submitted-object stack and
H-CRSPV receive identical artifacts and native evidence. The former accepts
6/6; H-CRSPV reconstructs the requirement set and rejects 6/6 before backend
dispatch (Table~\ref{tab:rq1}). Every row changes the deterministic output at
the attacked coordinates.

\begin{table}[!t]
\centering
\caption{Omission attacks with otherwise valid submitted evidence. All H-CRSPV
rows reject during exact-realization admission.}
\label{tab:rq1}
\TableFont
\begin{tabularx}{\columnwidth}{@{}Yrrr@{}}
\toprule
Omitted or stale relation & Coordinates & \shortstack{Submitted\\objects} &
H-CRSPV \\
\midrule
RangeOK & 1 & Accept & Reject \\
RoundDiv & 2 & Accept & Reject \\
Reindex & 8 & Accept & Reject \\
Stale predecessor & 1 & Accept & Reject \\
DARE LocalMap & 1 & Accept & Reject \\
Output artifact bind & 1 & Accept & Reject \\
\bottomrule
\end{tabularx}
\end{table}

\subsection{RQ2: Lifecycle and Finalization Safety}
The lifecycle matrix checks admission to \State{Provisional}, consumer rejection
before finalization, deadline ordering, valid and malformed fraud proofs,
replay, bond timing, terminal immutability, and idempotence. Each row compares
release, certificate, bond, AdmissionRecord, consumability, and application
state across four validators. Every control reaches its specified lifecycle
outcome, and no exact admission is counted as final acceptance. The frozen
counts remain in Appendix~\ref{app:matrix}.

\begin{table}[!t]
\centering
\caption{Lifecycle controls and four-validator outcomes.}
\label{tab:lifecycle}
\TableFont
\begin{tabularx}{\columnwidth}{@{}L{0.28\columnwidth}L{0.22\columnwidth}Y@{}}
\toprule
Control & Release result & Safety check \\
\midrule
Exact evidence & Provisional & No accepted record; not consumable \\
Valid timely challenge & Invalid & Certificate revoked; bond settled once \\
Malformed challenge & Unchanged & Rejected transaction; no advancement \\
Bad evidence & Rejected & One terminal settlement \\
Incomplete at deadline & Expired & Audit retained; bond available \\
No dispute at deadline & Active & $W$ preserved; atomic final record \\
Replay or mutation & Unchanged & No rollback or duplicate settlement \\
\bottomrule
\end{tabularx}
\end{table}

\subsection{RQ3: Dynamic Late Binding}
With one unchanged protocol authority, \FinalEOneLegal{}/\FinalEOneLegal{}
legal partitions accept, \FinalEOneInvalid{}/\FinalEOneInvalid{} malformed
partitions reject, and all 13 outcomes converge across four validators.

One continuous run uses a provenance-bound \VThirteenQwenModel{}
artifact~\cite{qwen2024qwen25}, four validators, and
\VThirteenProviderProcesses{} loopback HTTP provider processes on one host.
At $t_0$ the authority fixes the program, 240 local obligations, backend policy,
and $W=\VThirteenChallengeWindow{}$; only $r_1$ exists. The $r_2$ bundle appears
after its candidate commits and is withdrawn after finalization (HTTP 404
without changing terminal state); $r_3$ appears only afterward. No future
member list, object ID, root, or nonce exists at $t_0$. All
\VThirteenLegalStates{} releases become consumable; omission, overlap, stale,
and cross-binding controls reject; and all \VThirteenConvergedCases{} outcomes
agree across replicas.

A fixed concrete plan finds complete requirements only for $r_1$. The trusted
online planner and fixed semantic inspector find complete requirements for
$r_1/r_2/r_3$, while H-CRSPV additionally binds actual objects and finalizes all
three releases (Table~\ref{tab:late-binding}). The inspector establishes that
requirement completeness need not depend on an online trusted planner. Only the
H-CRSPV row owns the replicated transition to consumability. Across those
releases, H-CRSPV commits 11 aggregate 3-of-4 BLS certificates, accounting for
33 signer participations; the artifact does not claim 33 separately retained
partial-signature receipts.

\subsection{RQ4: Source Binding and Fair Baselines}
The restricted frontend processes two independently generated adapter pairs
under three fusion seeds. It binds all six source instances to their consumer
outputs and rejects 60/60 mutations of rank, alpha, orientation, target module,
tensor key, dtype, consumer, output mapping, or base identity.

The attack harness creates six omission families for each honest run, yielding
\FinalETwoAttacks{} loadable Safetensors artifacts. PEFT loads all artifacts,
and the submitted-object stack accepts 36/36 because every supplied object is
valid. H-CRSPV rejects 36/36 by reconstructing the omitted obligation. The
trusted canonical planner, deterministic replay, and fixed semantic inspector
also reject 36/36 while all four complete paths accept 6/6 honest releases.
Thus requirement completeness is achievable by several fair baselines.
H-CRSPV's additional responsibility is late-bound actual-object resolution and
the replicated provisional-to-consumable lifecycle.

For comparability, Table~\ref{tab:baselines} reports only local admission work
under single-process Python 3.10.9 with resident inputs: 30 repetitions for
each of six honest runs ($n=180$). The trusted-planner row includes canonical
plan construction and replay includes deterministic execution. Evidence
materialization and verification, retrieval/cache state, network transfer, and
four-validator consensus are excluded from every row.

\begin{table}[!t]
\centering
\caption{Fair-baseline decisions and local admission-only latency.}
\label{tab:baselines}
\TableFont
\begin{tabularx}{\columnwidth}{@{}L{0.25\columnwidth}Yrrr@{}}
\toprule
Acceptor & Completeness owner & \shortstack{Attacks\\rejected} &
\shortstack{Honest\\accepted} & \shortstack{Latency (ms)\\mean/P95} \\
\midrule
Submitted-object stack & Submitted objects & 0/36 & 6/6 & 0.30/0.43 \\
\system & Replicated exact set & 36/36 & 6/6 & 6.32/7.29 \\
Trusted canonical & Online planner & 36/36 & 6/6 & 5.00/5.69 \\
Full replay & Deterministic executor & 36/36 & 6/6 & 8.87/9.27 \\
Fixed semantic inspector & Fixed policy & 36/36 & 6/6 & 0.79/1.14 \\
\bottomrule
\end{tabularx}
\end{table}

On this small one-shot workload, full replay is faster end to end; evidence
materialization costs and model-output effects are reported separately in
Appendix~\ref{app:supp-eval}. H-CRSPV targets settings where validators cannot
all retrieve replay inputs, evidence is reused, or specialized evidence
backends and a shared release lifecycle are already required.

\subsection{RQ5: Replicated Artifact Admission}
One continuous chain binds a prerelease layout, static authorization, and one
immutable Safetensors digest through lowering, PEFT consumption, admission, and
the final record. The chain reproduces the same entry--group--object mapping as
RQ1 and uses a real 3-of-4 BLS aggregate generated by four centrally held test
keys. Across the fixed and dynamic controls, every committed decision and
application-state hash agrees on four validators. Detailed case inventories,
signer-isolation receipts, and state-transition counts are retained in
Appendix~\ref{app:matrix}.

\subsection{RQ6: Cost and Deployment Boundary}
Replay is simpler and faster for small available inputs. At the single
fresh-cache $r_2$ operating point, four validators retrieve
\VThirteenFreshReplayBytes{} replay bytes versus
\VThirteenFreshEvidenceBytes{} H-CRSPV bytes
(\VThirteenFreshByteRatio{}$\times$), or
\VThirteenFreshInclusiveRatio{}$\times$ after provider-plane traffic. A warm
observation records \VThirteenWarmReplayIncrementalBytes{}/%
\VThirteenWarmEvidenceIncrementalBytes{} incremental bytes. These are byte
measurements, not latency or same-assurance comparisons; the size-matched CPU
study in Appendix~\ref{app:cost} is a separate negative result.

The anonymous artifact binds every headline value to machine-readable results
and includes sources, localnet scripts, public projections, and a detached
clean-room verifier. Supporting test totals and claim-to-evidence mappings
appear in Appendices~\ref{app:validation-campaigns} and~\ref{app:matrix}.

\section{Discussion and Limitations}
\label{sec:discussion}
H-CRSPV targets late-bound releases whose evidence objects are resolved in
replicated state and whose consumers need ordered activation; replay remains
preferable when complete inputs are cheap. A fixed inspector closes
identifier-set omission but does not own grouping, object binding, or
activation. Claims remain conditional on Table~\ref{tab:guarantee-ladder};
Appendix~\ref{app:boundaries} states the full TCB and deployment limits.

\section{Related Work}
\label{sec:related}
Complete mediation and plan-to-execution systems motivate full event and plan
checking~\cite{saltzer1975protection,schneider2000enforceable,liu2025propertygpt,kim2025laputa,li2026ace}; provenance, transparency, verifiable-computation,
and translation systems authenticate workflows or supplied realizations
~\cite{torresarias2019intoto,slsa2023,newman2022sigstore,birkholz2026scitt,necula1997pcc,parno2013pinocchio,lycklama2024artemis,pnueli1998translation,leroy2009compcert}; model-composition and supply-chain work defines the artifact
substrate~\cite{hu2022lora,ilharco2023taskarithmetic,yadav2023ties,jiang2022modelsupply,wang2025mergebackdoor}. H-CRSPV instead owns required-set construction,
proposer-grouping policy, object resolution, and replicated activation
(Table~\ref{tab:boundary-ownership}); Appendix~\ref{app:related-details} gives
the full comparison.

\section{Conclusion}
H-CRSPV makes evidence completeness a replicated invariant: obligations map to
entries, proposer-induced groups pass policy and bind resolved objects, and
only guarded finalization activates a release. Explicit ownership preserves
late binding without redefining what must be proved.

\section*{Ethics Considerations}
The experiments use controlled model artifacts, local harnesses, and an isolated four-validator localnet; they do not attack production services or use human-subject data. The attack describes a class of verification-interface failures rather than a secret vulnerability in a named deployed system. We avoid presenting component proof systems, PEFT, provenance tools, Cosmos, or CometBFT as broken in their native threat models and will disclose any concrete implementation vulnerability discovered during integration to the affected maintainers before publication.

\IEEEtriggeratref{43}
\bibliographystyle{ieeetr}
\bibliography{references}

\section*{Generative-AI Use Disclosure}
OpenAI Codex desktop was used interactively in July 2026; the client did not expose a stable model or build identifier. It assisted with prose drafting, restructuring, and compression in the Abstract, Introduction, design and security-model sections, evaluation framing, Discussion, Related Work, Conclusion, tables, figure captions, and this disclosure. It also assisted with code scaffolding, test generation, experiment orchestration, result-packaging scripts, and diagnostic review. No model was permitted to manufacture measurements or silently replace missing results: the release source imports frozen machine-readable results and fails if its generated result macros are absent. The authors defined the research question and protocol, supplied and reviewed the implementation and source materials, executed or supervised the experiments, inspected raw logs, verified result-to-claim mappings and citations, and edited every generated passage and code change. The authors accept responsibility for the complete submission; no generative system is listed as an author.

\appendices

\section{Size-Matched Assurance--Cost Boundary}
\label{app:cost}
The small PEFT workload favors replay. We retain a separate \emph{Qwen2.5-3B-sized adapter-space workload} only as an operating-region study. Its package does not establish an originating repository revision, source Safetensors hashes, or lowering certificate and therefore is not a provenance-verified Qwen artifact. Its legal policy requires mandatory typed bindings, one deterministic KZG global relation, and exhaustive CertifiedLocal coverage; a $k=8$ opening diagnostic cannot discharge local obligations.

For validator count $V$, reuse $R$, bandwidth $B$, online evidence bytes $S_E$, and replay-input bytes $S_X$, the modeled accounting is
\begin{align*}
C_H&=T_{\mathrm{reg}}/R+V(T_{\mathrm{verify}}+S_E/B),\\
C_R&=V(T_{\mathrm{replay}}+S_X/B).
\end{align*}
The retained size-matched study derives 5,420 obligations (5,400
\textsc{CertifiedLocal}, 19 \textsc{Mandatory}, and one
\textsc{GlobalAlgebraic}) and measures an 18-realization historical operating
point. It is not used as evidence for the current A--D group/object or lifecycle claims.
Thirty Rust rows give 167.943/178.765~ms mean/P95 for the production-SRS global
relation and 2,663.263/2,729.535~ms for deterministic replay. Requirement
reconstruction and admission are single observed stages of 3,042.165 and
3,349.522~ms. Separately measured components model a policy-complete validator
at \FinalQwenValidatorSeconds{}~s versus \FinalQwenReplaySeconds{}~s for replay.
This is a negative CPU result,
not a crossover or observed end-to-end Qwen claim. Replay also removes the
CertifiedLocal quorum assumption, so the comparison is not same-assurance.

Separate planner tests accept canonical and cost-optimized realizations for 160 randomized legal configurations and reject 960 resealed omission, incompatible-batch, wrong-predecessor, downgrade, DA-expiry, and challenge-window attacks. The objective is synthetic and makes no optimality claim. A fresh-process scaling run accepts all 40 rows from $10^2$ to $10^5$ obligations; at 100,000 obligations, checker mean/P95 is 493.3/663.5~ms, total plan-plus-check mean/P95 is 588.0/805.8~ms, and maximum peak RSS is 142.0~MiB on the stated 8-CPU macOS host.

\section{Supplementary Evaluation Evidence}
\label{app:supp-eval}
The detailed dynamic late-binding comparison follows.

\begin{table}[!t]
\centering
\caption{Late-binding comparison. ``Yes'' under $r_i$ means that the mechanism
identifies a complete requirement set for that release.}
\label{tab:late-binding}
\TableFont
\setlength{\tabcolsep}{1.8pt}
\begin{tabularx}{\columnwidth}{@{}L{0.21\columnwidth}C{0.055\columnwidth}C{0.055\columnwidth}C{0.055\columnwidth}C{0.14\columnwidth}C{0.14\columnwidth}Y@{}}
\toprule
& \multicolumn{3}{c}{Complete requirements} & \shortstack{Actual\\object} & \shortstack{Replicated\\finality} & Added online actor \\
\cmidrule(lr){2-4}
Mechanism & $r_1$ & $r_2$ & $r_3$ & bound & owned & \\
\midrule
Frozen plan & Yes & No & No & No & No & None \\
Trusted planner & Yes & Yes & Yes & No & No & Planner call \\
Fixed inspector & Yes & Yes & Yes & No & No & None \\
H-CRSPV & Yes & Yes & Yes & Yes & Yes & Certifier quorum \\
\bottomrule
\end{tabularx}
\end{table}

The following tables retain the measurements needed to interpret the main-text
claims. Full identities, raw receipts, rejection logs, and the claim lock remain
in the frozen artifact. For E3, each lifecycle is one observation ($n=1$);
timing stages are disjoint, and the byte ledger separates cache hits from
reusable and release-bound bytes.

\paragraph{Deferred audit details}
The production profile identifier is \texttt{relationir-core-v1}; its full
SHA-256 value (abbreviated \texttt{b0af034e}\ldots\texttt{893f70e4}) is bound into program,
obligation, and transcript identities. The authenticated co-reference gate
accepts the honest registrar transcript and rejects 18/18 context, descriptor,
root, domain, length, and canonical-stream substitutions. One prerelease
in-toto layout and static workflow authorization bind a single Safetensors
digest before any release; unsigned proposer envelopes later select release
realizations, while release-bound aggregate certificates remain required.

The isolation launcher records \VThirteenIsolationReceipts{} receipts,
\VThirteenPrivateCanarySuccesses{} successful private-key canary reads,
\VThirteenPrivateKeyMounts{} private-key mounts, and
\VThirteenSignerSockets{} signer sockets after signer exit. A separate offline
Qwen control retains \VThirteenBehaviorSamples{} samples and the corresponding
native and omission receipts. Supporting campaigns accept 128/128 generated
legal manifests and 90/90 honest plans; reject 466/466 requirement mutations
and 90/90 concrete attacks; report no Python/Rust/Go mismatch over 10,000 legal
programs; expose all 12 targeted guard ablations; and detect all 15 injected
failures in 2,729 local-judge cases. These records support implementation
auditability and do not alter the main RQ outcomes.

\subsection{Retrieval and Lifecycle Accounting}
Table~\ref{tab:byte-ledger} counts validator-side HTTP response bodies. In the
fresh $r_2$ row, \VThirteenFreshReusableBytes{} reusable-component bytes are
not cache hits: the measured cache-hit count is
\VThirteenFreshCacheHitBytes{}. Release-bound traffic is
\VThirteenFreshNewBytes{} bytes across four validators, i.e.,
\VThirteenFreshPerValidatorNewBytes{} unique source bytes fetched once per
validator. The table excludes \VThirteenProviderPlaneBytes{} bytes of
provider-plane traffic; RQ6's inclusive ratio adds that traffic explicitly.

\begin{center}
\begin{minipage}{\columnwidth}
\centering
\captionof{table}{E3 validator-retrieval byte ledger (aggregate HTTP response bodies).}
\label{tab:byte-ledger}
\TableFont
\begin{tabularx}{\columnwidth}{@{}Yrrr@{}}
\toprule
Metric (bytes) & \shortstack{Fresh\\$r_2$} &
\shortstack{Continuous\\$r_1\!\rightarrow\!r_2\!\rightarrow\!r_3$} &
\shortstack{Warm\\$r_1\!\rightarrow\!r_2$} \\
\midrule
Replay input & 12,396,243,912 & -- & 0 \\
H-CRSPV retrieval & 10,146,896 & 11,140,644 & 506,996 \\
Reusable component & 9,639,900 & -- & 9,639,900 \\
Cache hit & 0 & 19,279,800 & 9,639,900 \\
Release-bound & 506,996 & 1,500,744 & 506,996 \\
Stale response & 0 & 953,252 & 446,256 \\
\bottomrule
\end{tabularx}
\end{minipage}
\end{center}

Each release is one wall-clock observation ($n=1$). In
Table~\ref{tab:e2-lifecycle-timing}, \emph{Gap} is measured wall time between
instrumented stages, and \emph{Residual} is end-to-end time minus the sum of
all named stages and Gap. A zero residual is an accounting identity, not a
variance or accuracy estimate.

\begin{center}
\begin{minipage}{\columnwidth}
\centering
\captionof{table}{E3 wall-clock lifecycle decomposition (ms); admission and finalization stages are disjoint.}
\label{tab:e2-lifecycle-timing}
\TableFont
\begin{tabularx}{\columnwidth}{@{}Yrrr@{}}
\toprule
Stage & $r_1$ & $r_2$ & $r_3$ \\
\midrule
Candidate & 5,411.49 & 3,109.96 & 1,993.44 \\
Publish & 5,805.65 & 7,730.45 & 9,669.22 \\
Catalog & 23.57 & 16.75 & 20.97 \\
Plan & 42.21 & 31.51 & 35.31 \\
Materialize+BLS & 21,962.55 & 86,873.13 & 135,517.20 \\
Admit & 8,047.30 & 8,130.41 & 7,781.96 \\
Policy wait & 239,540.68 & 169,596.33 & 124,102.82 \\
Finalizer & 615.89 & 1,612.12 & 635.73 \\
Gap & 360.43 & 350.42 & 455.19 \\
\midrule
End-to-end & 281,809.77 & 277,451.08 & 280,211.84 \\
Residual & 0.00 & 0.00 & 0.00 \\
\bottomrule
\end{tabularx}
\end{minipage}
\end{center}

\subsection{Attack Controls and Secondary Evidence}
For every attack in Table~\ref{tab:e2-attacks}, a nonzero evidence-transaction
ABCI code means deterministic rejection with no admitted release mutation; the
subsequent cancel transaction returns code zero, meaning that cleanup commits
successfully. The packaged verifier recomputes the owning gate from the raw log
rather than trusting the normalized label.

\begin{center}
\begin{minipage}{\columnwidth}
\centering
\captionof{table}{Committed E3 attack controls and independently recomputed rejection gates.}
\label{tab:e2-attacks}
\TableFont
\begin{tabularx}{\columnwidth}{@{}L{0.14\columnwidth}Y C{0.20\columnwidth}C{0.17\columnwidth}@{}}
\toprule
Attack & Mutation; owning gate & \shortstack{Evidence /\\cancel ABCI} & Replicas \\
\midrule
Omission & Remove one required entry; entry/group exactness & $>0$ / $0$ & Reject; 4/4 \\
Overlap & Cover one member twice; typed disjointness & $>0$ / $0$ & Reject; 4/4 \\
Stale & Reuse a prior-release payload; release/object binding & $>0$ / $0$ & Reject; 4/4 \\
Cross-binding & Bind another program or artifact; artifact/program binding & $>0$ / $0$ & Reject; 4/4 \\
\bottomrule
\end{tabularx}
\end{minipage}
\end{center}

Table~\ref{tab:secondary-eval} separates output-level consequences from
local-judge conformance. The 36 artifacts are the RQ4 omission artifacts; the
2,729 judge cases and 15 injected failures are a distinct implementation
campaign and are not added to the attack denominator.

\section{RelationIR-Core v1 Requirement Rules}
\label{app:rules}
Table~\ref{tab:rules} summarizes the rule families used by the current checker. A future extension is covered only after it adds a type rule, base obligations, propagation conditions, normalized keys, admissible backends, and a witness or interpreter interface.

\begin{table*}[!t]
\centering
\caption{Requirement rules implemented by the closed \Pzero profile.}
\label{tab:rules}
\TableFont
\begin{tabularx}{\textwidth}{@{}L{0.13\textwidth}L{0.23\textwidth}L{0.25\textwidth}Y@{}}
\toprule
Primitive & Base requirements & Propagated requirements & Evidence path \\
\midrule
Input/output descriptors & Artifact, type, domain, commitment, reachability & Output binding and lineage & Mandatory typed checks \\
AffineCombine & Coefficients, predecessors, affine relation & Dependency and commitment domains & Production KZG \\
LocalMap: drop/scale & Transform, seed/mask or scale, block witness & RangeOK and exact source & Direct or CertifiedLocal judge \\
LocalMap: round-div & Divisor, signed lift, tie rule & RangeOK and overflow & Direct or CertifiedLocal judge \\
LocalMap: quantize & Scale, clipping, cast, rounding & Signed input/output ranges & Direct or CertifiedLocal judge \\
Reindex & Total block map; unique outputs & Layout and dependency mapping & Deterministic judge \\
Provisional local backend & Relation digest and signer set & DA, challenge, fraud, finalization & BLS certificate and lifecycle \\
\bottomrule
\end{tabularx}
\end{table*}

\FloatBarrier

\subsection{Normative Integer and Tensor Semantics}
The current \Pzero language evaluates canonical integer tensors, not host floating point. A tensor type
$T=(\mathit{shape},\mathit{layout},w,s,B,q,d)$ binds shape and coordinate order, signed width $w$, fixed-point scale $s$, declared absolute bound $B$, field modulus $q$, and commitment domain $d$. A decoded coordinate $x$ is valid only when
\begin{equation}
-2^{w-1}\leq x<2^{w-1},\qquad |x|\leq B<q/2,
\end{equation}
and its field encoding is $x\bmod q$. The $B<q/2$ premise makes the registered signed lift unique; an affine field equality without the corresponding range obligations is insufficient.

For $d>0$, the only registered division rule is nearest-integer rounding with halves away from zero:
\begin{equation}
\mathsf{rnd}_d(x)=\mathsf{sgn}(x)
  \left\lfloor\frac{|x|+\lfloor d/2\rfloor}{d}\right\rfloor,
\end{equation}
with $\mathsf{rnd}_d(0)=0$. The implemented primitives have the following coordinate semantics, with every displayed parameter loaded from the registered node rather than from a witness:
\begin{align*}
\mathsf{Scale}_{a}(x)_j &= a x_j,\\
\mathsf{DropRescale}_{\sigma,a}(x)_j
  &= \begin{cases}a x_j&\mathsf{Keep}(\sigma,j)=1,\\0&\text{otherwise},\end{cases}\\
\mathsf{Affine}_{\boldsymbol a}(x_1,\ldots,x_m)_j
  &=\sum_{i=1}^{m}a_i x_{i,j},\\
\mathsf{RoundDiv}_{d}(x)_j&=\mathsf{rnd}_d(x_j),\\
\mathsf{Quantize}_{d,\ell,u}(x)_j
  &=\min(u,\max(\ell,\mathsf{rnd}_d(x_j))),\\
\mathsf{Reindex}_{\pi}(x)_j&=x_{\pi(j)}.
\end{align*}
$\mathsf{Keep}$ is a versioned deterministic bit generator bound to seed $\sigma$ and global coordinate $j$. The affine range rule requires a declared output bound at least $\sum_i|a_i|B_i$ and below $q/2$; scaling, rounding, and quantization propagate the corresponding arithmetic bounds. $\pi$ must be a bijection over $[0,n)$; the implemented packing profile additionally requires each output block, including a final partial block, to select one contiguous, aligned source block. These predicates define the current claim; alternative stochastic masks, zero-point quantization, arbitrary reshapes, and host-language casts are rejected as unsupported.

\subsection{Normative Canonicalization}
Protocol identities hash a versioned, domain-separated canonical encoding. Accepted structured values use sorted object keys, order-preserving arrays, base-10 integers only, UTF-8 text with deterministic escaping, and no insignificant whitespace; mathematical sets are sorted by their canonical byte encoding before hashing. Tensor bytes use registered signed width, endianness, coordinate order, length, and block partition. The registration, obligation, witness, transcript, and output domains use distinct tags. A production decoder must reject duplicate identifiers, noncanonical numeric forms, unknown primitive versions, and any descriptor/byte-length mismatch before computing an identity. The independent Rust oracle tests the current Python encoding, but neither implementation is mechanized; this bridge remains in the TCB.

\section{Extended Scope and Deployment Boundaries}
\label{app:boundaries}
The main result treats RelationIR as authoritative. For the supported declarative
PEFT/Safetensors subset, \textsf{CheckLower} binds source configuration, tensor
bytes, consumer mapping, and output identity to \Pzero; 60 source/IR mutations
exercise this bridge. A separate checker validates a 71-event executor trace and
rejects 9/9 mutations, including four that recompute the program hash after
changing source-relevant fields. The executor parses PEFT-format artifacts but
does not instrument arbitrary PEFT, Python, Torch, or custom kernels; unsupported
execution is rejected.

Exact binding moves security weight onto decoding and canonicalization. Ambiguity in signed encodings, endianness, layout, shard order, or duplicate identifiers remains a file-to-vector TCB failure rather than a backend-proof failure. CertifiedLocal soundness and finalization remain conditional on the configured committee, data availability, watcher, challenge inclusion, and deadline assumptions; deployments that reject these assumptions require a liveness-independent proof or replay for critical relations. Finally, the four-validator localnet demonstrates current-schema admission and deterministic replicated rejection, not sustained throughput, external-DA retention, or production-chain finality. A centralized auditor can run the same checker; consensus is useful only when multiple parties require shared ordering and consequences.

\section{Extended Related Work}
\label{app:related-details}

H-CRSPV composes established primitives around a distinct enforcement
boundary. The relevant comparison is whether a system owns required-set
construction, policy validation of proposer grouping, concrete-object
resolution, and activation when
a later actor controls realization (Table~\ref{tab:boundary-ownership}).

\paragraph{Reference monitors and complete mediation}
Complete mediation checks every security-relevant access, whereas execution
monitors enforce policies recognizable from observed events
~\cite{saltzer1975protection,schneider2000enforceable}. H-CRSPV specializes this
principle to plan-to-evidence admission: it derives the reference set before
realization and mediates entries, proposed groups, resolved objects, and final
activation.

\paragraph{Specification and plan-to-execution enforcement}
PropertyGPT generates and checks candidate smart-contract properties
~\cite{liu2025propertygpt}; LAPUTA checks Spark physical plans and protects
their execution~\cite{kim2025laputa}; ACE separates trusted abstract plans from
untrusted concrete mappings~\cite{li2026ace}. H-CRSPV starts from an
authenticated finite requirement set and extends plan-to-execution enforcement
through registered-policy validation of proposed evidence sharing,
keeper-object binding, and the provisional-to-active transition.

\paragraph{Supply-chain provenance and transparency}
in-toto and SLSA bind declared steps and provenance, while Macaron evaluates
logic-based supply-chain policies
~\cite{torresarias2019intoto,slsa2023,hassanshahi2023macaron}.
Sigstore/Rekor, TUF, Uptane, Policy Transparency/PolyLog, SCITT, and Scudo add
signing, update, transparency, receipt, and root-composition layers
~\cite{newman2022sigstore,samuel2010tuf,kuppusamy2018uptane,ferraiuolo2022policytransparency,birkholz2026scitt,moore2024scudo}.
These systems authenticate workflow assertions and their history. They do not
automatically derive the complete semantic-relation multiset of a late-bound
model release or bind it to one replicated consumer state; they can enforce
H-CRSPV's boundary when policy also supplies the constructor, object resolver,
and final-state rule.

\paragraph{Proof-carrying and verifiable computation}
Proof-carrying and verifiable-computation systems check supplied executions or
claims~\cite{necula1997pcc,chiesa2010pcd,parno2013pinocchio,weng2021mystique,liu2021zkcnn,lycklama2024artemis,sun2023zkdl,sun2024zkllm,wang2023zkfl,liao2026verilora};
under-constrained-circuit analyses find unintended witnesses
~\cite{pailoor2023underconstrained,chaliasos2024snarksok}.\@ \sove instead omits
an externally required statement: every received proof may be sound while one
required proof is absent. A proof system closes that gap only when its public
statement, circuit, or admission layer also commits to required-set equality.

\paragraph{Compilation and optimistic verification}
Translation validation, CompCert, and Tenspiler connect programs to checked
realizations~\cite{pnueli1998translation,leroy2009compcert,qiu2024tenspiler};
fraud-proof and outsourced-computation systems challenge available invalid
transitions or replicated results
~\cite{kalodner2018arbitrum,albassam2018fraudda,zheng2021agatha,das2019yoda}. H-CRSPV's
frontend is narrower: it binds one declarative PEFT/Safetensors subset to a
closed RelationIR profile, then keeps every bound relation provisional while
its watcher-completeness premise applies.

\paragraph{Model-artifact composition}
LoRA, task arithmetic, sparse merging, and adapter composition define the
composition substrate
~\cite{hu2022lora,ilharco2023taskarithmetic,yadav2023ties,yu2024supermario}.
Model-supply-chain studies, merge-time backdoors, Model DNA, and Atlas motivate
pipeline-wide integrity and lineage
~\cite{jiang2022modelsupply,meiklejohn2025mlsupply,wang2025mergebackdoor,mu2023modeldna,spoczynski2025atlas}.
These works expose or track unsafe components and compositions; H-CRSPV instead
enforces release-time completeness for one closed, authority-declared profile.

\section{Proof Details}
\label{app:proofs}
\subsection{Finite Requirement Construction}
For a fixed well-typed program and block partition, registered nodes, edges,
blocks, outputs, and policy clauses are finite. Equation~\eqref{eq:reqderive}
therefore yields a finite list, and canonical sorting terminates. The
implementation uses topological enumeration plus validation, not a fixed-point
or worklist algorithm.

\subsection{Replica Determinism}
Fix canonical bytes for $\rho$, $\Gamma$, $\mathcal E$, and the authenticated
keeper snapshot $\mathcal K$. Canonical decoding either fails at every correct
replica or yields the same typed values. On the accepting branch,
$\Req(\rho)$ is a pure topological traversal of the same finite DAG;
$\Norm(\Gamma)$ is a pure typed normalization; canonical sorting followed by a
duplicate ordinal fixes every occurrence tag; and equality of
$\kappa_\rho$ fixes the same proposer-induced quotient classes. Each class and
object key is
then traversed in canonical byte order. Finally, the resolver is a function of
the same release-scoped payload and keeper snapshot. Functional composition
therefore gives identical $R$, $M$, $G$, $\mathcal O_\rho$, and acceptance
bits. This argument requires the canonical decoder and resolver assumptions in
Table~\ref{tab:guarantee-ladder}; it is not a claim about faulty replicas.

\subsection{Lemma 1}
Let $R=\Req(\rho)$ and $M=\Norm(\Gamma)$. Registration freezes $\rho$.
For each $o\in\mathcal U_\rho$, the checker counts the occurrence-tagged entries
whose typed projection is $o$. Acceptance establishes
$\mathsf{mult}_{(\pi^\rho_{\rm obl})_\#\mathsf{Occ}(M)}(o)=\mathsf{mult}_R(o)$, including zero
on both sides for identities absent from $R$. Hence no occurrence can project to
an unexpected identity, and every required occurrence has exactly one matching
entry occurrence. The remaining accepting branches compare each entry's
authority hash, node, relation, primitive, endpoints, coordinates, parameters,
dependencies, assurance policy, and interpreter version with the obligation to
which it projects. They establish
$\mathsf{ExactEntryBind}_\rho(m,\pi^\rho_{\rm obl}(m))$ for every
$m\in\mathsf{Occ}(M)$. These two facts are exactly
Eq.~\eqref{eq:entry-complete}; equivalently, they induce a
multiplicity-preserving bijection between required and admitted occurrences.

\subsection{Lemma 2}
Equality of the complete typed key $\kappa_\rho$ induces the equivalence
relation $\sim_{\rm mat}$, so the grouping proposed by those entry keys has a
unique quotient whose classes partition $\mathsf{Occ}(M)$. These are proposed
groups, including any singleton classes. Since
$O_g=(\pi^\rho_{\rm obl})_\#g$ and pushforward
commutes with disjoint multiset union,
\begin{equation*}
\biguplus_{g\in G}O_g
=(\pi^\rho_{\rm obl})_\#\!\left(\biguplus_{g\in G}g\right)
=(\pi^\rho_{\rm obl})_\#\mathsf{Occ}(M)=R,
\end{equation*}
where the last equality is Lemma~1. The registered batching checks then
establish \textsf{GroupOK} for every proposed group; only those passing classes
are admissible, and non-shareable or incompatible classes reject. This check
does not prove that an object covers $O_g$. After all groups are admissible, the
resolver reads typed payloads and registered state, not a proposer-supplied
digest, and accepts exactly one object per class only if both its canonical
statement and full materialization key agree. Totality of $\mu$ excludes a
group with no object, injectivity excludes cross-group object reuse, and
surjectivity excludes any resolved orphan object outside the image of a group.
The two equality checks reject cross-context or unrelated statements. This proves
Eqs.~\eqref{eq:materialization-complete} and~\eqref{eq:object-bind}; constructor
or resolver bugs remain explicit TCB failures.

\subsection{Proof of \ExactRealizationTheorem{}}
The structural accepting path is the sequential composition of the entry phase
and materialization/object phase over one authenticated authority snapshot.
Lemma~1 supplies the exact obligation--entry multiset equality. Lemma~2
reconstructs the quotient proposed by complete entry keys, checks its disjoint
complete union, admits every group under \textsf{GroupOK}, resolves the actual
typed objects from payloads and keeper state, and establishes the
canonical-statement bijection. Their conjunction is $\StructOK_\rho$,
equivalently the negation of Eq.~\eqref{eq:realization-gap}, and establishes
Eq.~\eqref{eq:exact-realization}. No step invokes a backend truth-lifting rule,
source-lowering claim, external co-reference adequacy premise, watcher, or
finalizer; those premises enter only the semantic and lifecycle results.

\subsection{Operational--Abstract Correspondence}
Fix one operational instance
$\OpRep_{\rho,\ell}(q,\mathcal K)=(\Gamma,\mathcal E,A_{\mathrm{out}})$.
On the accepting branch, Lemma~2 yields the unique bijection
$\mu:G\xrightarrow{\sim}\mathcal O_\rho(\mathcal E,\mathcal K)$. We therefore
instantiate $\SubObj(q)$ by its codomain and define
$\Cov_{\rho,\ell}(q,\mu(g))=O_g$.

This definition is total on $\SubObj(q)$ because $\mu$ is surjective. It is
single-valued because $\mu$ is injective and the resolver rejects duplicate
complete object keys. It is authority-scoped because $\mathsf{GroupOK}$ checks
the exact obligation multiset and both equalities in
Eq.~\eqref{eq:object-bind} bind the canonical statement, complete context, and
materialization key. Finally,
\begin{equation*}
\begin{aligned}
\biguplus_{e\in\SubObj(q)}\Cov_{\rho,\ell}(q,e)
 &=\biguplus_{g\in G}\Cov_{\rho,\ell}(q,\mu(g))\\
 &=\biguplus_{g\in G}O_g=R_\rho,
\end{aligned}
\end{equation*}
where the first equality reindexes by the bijection and the last is
Eq.~\eqref{eq:materialization-complete}. Thus both clauses of
Eq.~\eqref{eq:complete-realization} hold. Conversely, contraposition shows that
an abstract SOVE accepted by a submitted-object interface on this operational
instance must expose at least one operational realization-gap branch; no reverse
equivalence is needed because the operational checker may reject additional
malformed or unsupported realizations.

\subsection{Lemma 3}
Let $s_0,\ldots,s_k=s$ witness $\Reach_\rho(s)$. We prove
Eq.~\eqref{eq:reachable-finalization} by induction on $k$. At $k=0$, the initial
state contains neither \texttt{ACTIVE} nor an accepted record, so the antecedent
is false. For the inductive step, consider
$(s_{k-1},a_{k-1},s_k)\in\Delta_\rho$. If the release was already consumable in
$s_{k-1}$, the induction hypothesis supplies a witnessing finalizer; terminal
immutability preserves exact admission, zero disputes, certificate state, and
settlement in $s_k$. Otherwise, consumability is newly true in $s_k$. By the
writer restriction on $\Delta_\rho$, $a_{k-1}$ must be
$\mathsf{finalize}(\ell)$. Its guard supplies the immutable exact-admission
record, $h(s_{k-1})>d_\ell$, the preserved window, all-group eligibility, and
zero disputes, while its atomic write supplies active, accepted, finalized
certificates, and one-time bond settlement. Taking $u=s_{k-1}$ and $v=s_k$
witnesses $\mathsf{Finalized}_\rho(\ell,s_k)$.

No ordering corner case adds another branch: a challenge is enabled through
$h=d_\ell$, whereas finalization requires $h>d_\ell$; accepted transactions are
linearized within each block. Thus an included deadline challenge reaches
terminal \State{Invalid} before any later finalizer can be enabled. Rejection,
expiry, and upheld-challenge states are terminal and non-consumable, and replay
cannot repeat the atomic settlement.

\subsection{Proof of \SemanticCompositionProposition{}}
\SemanticCompositionProposition{} is conditional on primitive adequacy,
backend lifting, canonicalization, co-reference, and watcher completeness over
the registered \Pzero\ program. Lemmas~1--2 supply exactly the obligations and
actual objects used by Eq.~\eqref{eq:lifting}. Direct and KZG acceptance supplies
their typed semantic predicates by backend soundness. For a CertifiedLocal
class, assume for contradiction that its registered conjunction is false.
$\mathsf{WatcherComplete}_\rho(\ell)$ supplies detection, valid typed-challenge
construction, timely submission, and inclusion; judge completeness upholds the
challenge and transitions the release to terminal \State{Invalid}. This
contradicts reachable consumability by Lemma~3, so every class conjunction used
by a consumable release is true.

Now induct over the registered topological order. Registered input nodes form
the base case by the authority and co-reference premises. For node $n_j$, the
induction hypothesis equates every predecessor value with its reference value;
the exact \textsf{NodeReq} predicates and the stated per-node adequacy implication
then yield $V_j=V_j^*$. The output-map obligation and object binding identify the
final reference value with the consumer artifact. The present mixed KZG/Merkle
path still trusts registration to resolve both roots to one canonical stream.

\subsection{Scope of \SourceTransferClaim{}}
For the supported frontend, \textsf{CheckLower} reparses source configuration and
Safetensors metadata, recomputes content hashes, and compares every registered
source tensor, LoRA parameter, consumer port, and output mapping with the fields
used to construct $P$. Those checks establish recorded field equality, not a
general semantic theorem for Python. Observational equality of the supported
source evaluator and \Pzero\ evaluator is therefore named explicitly as
$\mathsf{FrontendRel}_\rho$ and kept in the frontend TCB. Conditional on that
invariant, \SemanticCompositionProposition{} transfers immediately to the bound
source output. Parser or correspondence bugs remain lowering-TCB failures.

\subsection{Corollary 1}
Any omitted or duplicated entry contradicts Lemma~1. Any overlapping,
incomplete, incompatible, reused, cross-bound, or orphan materialization
contradicts Lemma~2 and therefore \ExactRealizationTheorem{}. In every reachable
state, a release that remains provisional, has an open dispute, lacks exact
admission, or is marked settled without a witnessing finalizer cannot satisfy
Eq.~\eqref{eq:final-accept}. A provisional release may satisfy
Eq.~\eqref{eq:finalize-guard} only after the deadline with exact admission, all
groups eligible, and no open dispute; the enabled finalizer then atomically
changes it to \State{Active} with an accepted record. Hence no structurally
defective or unfinalized release is consumable, regardless of the validity of
the objects that were submitted.

\begin{center}
\begin{minipage}{\columnwidth}
\centering
\captionof{table}{Secondary evaluation evidence: output effects and local-judge conformance.}
\label{tab:secondary-eval}
\TableFont
\begin{minipage}[t]{\columnwidth}
\centering
\textit{(a) Effect distribution across 36 PEFT attack artifacts}\par\vspace{2pt}
\begin{tabular}{@{}lrrr@{}}
\toprule
Metric & Median & P95 & Max \\
\midrule
Logit $L_\infty$ & 0.0478 & 1.2254 & 2.2688 \\
Changed coordinates (\%) & 0.0244 & 1.4771 & 1.5381 \\
Weight-delta $L_2$ & 0.0560 & 1.4277 & 1.4940 \\
Top-1 position change (\%) & 0 & 6.25 & 16.67 \\
\bottomrule
\end{tabular}
\end{minipage}

\vspace{5pt}
\begin{minipage}[t]{\columnwidth}
\centering
\textit{(b) Local-judge conformance}\par\vspace{2pt}
\begin{tabular}{@{}lrr@{}}
\toprule
Primitive & Cases & \shortstack{Injected\\failures caught} \\
\midrule
Drop-rescale & 545 & 3/3 \\
RoundDiv & 551 & 3/3 \\
Quantize & 551 & 3/3 \\
RangeOK & 543 & 3/3 \\
Reindex/sharding & 539 & 3/3 \\
\midrule
Total & 2,729 & 15/15 \\
\bottomrule
\end{tabular}
\end{minipage}
\end{minipage}
\end{center}

\FloatBarrier

\section{Evidence Summary}
\label{app:matrix}
Table~\ref{tab:matrix} maps RQ1--RQ6 to frozen evidence and claim boundaries;
the full inventory, commands, hashes, and claim lock remain in the artifact.
The 39 structured lifecycle records decompose as
$1$ legal $+35$ committed attacks $+1$ replay $+2$ prepublication controls.
Accordingly, the 37 committed Cosmos decisions are the legal release, 35
attacks, and replay; ``37 attack controls'' elsewhere means the 35 committed
attacks plus the two prepublication controls.

\begin{table*}[!t]
\centering
\caption{Frozen evidence-to-claim summary. Counts with different denominators are not combined.}
\label{tab:matrix}
\TableFont
\setlength{\tabcolsep}{2.4pt}
\begin{tabularx}{\textwidth}{@{}C{0.045\textwidth}L{0.58\textwidth}Y@{}}
\toprule
RQ & Frozen observation & Claim boundary \\
\midrule
1 & \textbf{Exact entries/groups/objects.} \FinalRealizationEntries{} entries $\rightarrow$ \FinalMaterializationGroups{} admissible shared-evidence groups $\rightarrow$ \FinalActualEvidenceObjects{} objects; six omissions and 18 co-reference substitutions reject. & Trusted registrar; no KZG--Merkle equality proof; entries are not independent proofs. \\
2 & \textbf{Reachable lifecycle safety.} 37/37 committed Cosmos decisions converge: one legal release, \FinalChainAttacks{} attacks, and one replay; two prepublication controls remain separate. & Four-validator localnet, loopback DA, local finality. \\
3 & \textbf{Dynamic late binding.} E1 has \FinalEOneLegal{} legal and \FinalEOneInvalid{} invalid partitions with 13/13 convergence. In E3, the frozen plan is complete only for $r_1$; the trusted planner, fixed inspector, and H-CRSPV are complete for $r_1,r_2,r_3$. & Six same-host HTTP processes; each E3 lifecycle has $n=1$. \\
4 & \textbf{Source binding/fair completeness.} Six source instances bind; 60/60 source mutations and 36/36 loadable omissions reject; complete baselines accept 6/6 honest cases. & Declarative PEFT/Safetensors subset; frontend relation and interpreter remain in the TCB. \\
5 & \textbf{Replicated admission.} One artifact identity, prerelease authorization, 11 aggregate 3-of-4 certificates, and 33 signer participations retain one lineage. & Centralized test keys; unsigned release envelopes. \\
6 & \textbf{Cost/deployment boundary.} Fresh: \VThirteenFreshByteRatio{}$\times$ replay/evidence; warm: \VThirteenWarmReplayIncrementalBytes{}/\VThirteenWarmEvidenceIncrementalBytes{} bytes; size-matched CPU result is negative. & One fresh and one warm point; not latency, optimized replay, throughput, or same assurance. \\
\bottomrule
\end{tabularx}
\end{table*}

\FloatBarrier

\section{Supplementary Validation Campaigns}
\label{app:validation-campaigns}
\subsection{Requirement Construction}
Generated \Pzero\ tests accept 128/128 legal manifests and reject 466/466
mutations across 21 rule families; a multi-seed campaign accepts 90/90 honest
plans and rejects 90/90 endpoint-changing attacks. Python, Rust, and Go agree on
10,000 legal obligation roots, reject nine malformed cases and 30/30 profile
mutations, and agree on 1,882 bounded semantic cases. An independent Rust oracle
matches 10,000 valid programs, rejects 2,000 invalid programs, and direct
mutation rejects 50/50 non-equivalent changes across 18 families. These finite
campaigns reduce correlated implementation risk but do not prove profile or
rule-table adequacy.

\subsection{Guards and Local Semantics}
Thirty-two context, root, endpoint, coordinate, parameter, and reindex controls
match expectations. Twelve guard ablations cover anti-downgrade, batch
compatibility, DA expiry, and challenge timing: every attack rejects under the
full verifier and passes only its matching ablation; honest input passes both.
This shows necessity in the evaluated cases, not minimality. A separate
2,729-case judge harness detects all 15 injected failures. Because fixtures are
shared, this is conformance evidence, not formal validation.

\end{document}

%% file: v12_1_results_macros.tex
\renewcommand{\FinalChainAttacks}{35}

\renewcommand{\FinalRealizationEntries}{345}
\renewcommand{\FinalMaterializationGroups}{26}
\renewcommand{\FinalActualEvidenceObjects}{26}
\renewcommand{\FinalCertifiedLocalEntries}{320}

\renewcommand{\FinalEOneLegal}{3}
\renewcommand{\FinalEOneInvalid}{10}

\renewcommand{\FinalETwoAttacks}{36}

\renewcommand{\FinalAlternateDigests}{5}
\renewcommand{\FinalQwenValidatorSeconds}{7.870}
\renewcommand{\FinalQwenReplaySeconds}{5.582}

\renewcommand{\FinalWrongDigestResult}{\FinalAlternateDigests/\FinalAlternateDigests{} alternate digests self-verify and fail the keeper digest}

%% file: v13_results_macros.tex
\renewcommand{\VThirteenQwenModel}{Qwen2.5-1.5B}

\renewcommand{\VThirteenLegalStates}{3}

\renewcommand{\VThirteenConvergedCases}{7}
\renewcommand{\VThirteenChallengeWindow}{10}

\renewcommand{\VThirteenFreshReplayBytes}{12,396,243,912}
\renewcommand{\VThirteenFreshEvidenceBytes}{10,146,896}
\renewcommand{\VThirteenFreshByteRatio}{1,221.68}
\renewcommand{\VThirteenProviderPlaneBytes}{422,127}

\renewcommand{\VThirteenFreshInclusiveRatio}{1,172.88}
\renewcommand{\VThirteenFreshReusableBytes}{9,639,900}
\renewcommand{\VThirteenFreshCacheHitBytes}{0}
\renewcommand{\VThirteenFreshNewBytes}{506,996}
\renewcommand{\VThirteenFreshPerValidatorNewBytes}{126,749}

\renewcommand{\VThirteenWarmReplayIncrementalBytes}{0}
\renewcommand{\VThirteenWarmEvidenceIncrementalBytes}{506,996}

\renewcommand{\VThirteenBehaviorSamples}{24}

\renewcommand{\VThirteenIsolationReceipts}{3}
\renewcommand{\VThirteenPrivateCanarySuccesses}{0}
\renewcommand{\VThirteenPrivateKeyMounts}{0}
\renewcommand{\VThirteenSignerSockets}{0}
\renewcommand{\VThirteenProviderProcesses}{6}

%% file: authors.tex
\author{%
\IEEEauthorblockN{Weijie Miao\IEEEauthorrefmark{1}, Henry Hong-Ning Dai\IEEEauthorrefmark{2}, and Ming Li\IEEEauthorrefmark{1}}%
\IEEEauthorblockA{\IEEEauthorrefmark{1}Department of Industrial and Systems Engineering, The Hong Kong Polytechnic University, Hong Kong\\
\IEEEauthorrefmark{2}Department of Computer Science, Hong Kong Baptist University, Hong Kong\\
Corresponding author: Ming Li}%
}
\hypersetup{%
  pdftitle={H-CRSPV: Preventing Semantic Omission in Late-Bound Large Language Model Releases},%
  pdfauthor={Weijie Miao, Henry Hong-Ning Dai, Ming Li}%
}

%% file: figures/sove_attack_tikz.tex
\begin{tikzpicture}[
  x=1cm,y=1cm,
  box/.style={draw=black!75, line width=0.55pt, fill=black!3,
              align=center, inner sep=3.5pt, minimum height=0.86cm,
              font=\footnotesize},
  key/.style={box, fill=black!10, font=\footnotesize\bfseries},
  note/.style={box, dashed, fill=white},
  flow/.style={-{Stealth[length=1.7mm,width=1.2mm]}, line width=0.55pt},
  every node/.style={text depth=0.15ex}
]
  \node[font=\footnotesize\bfseries,anchor=west] at (0.15,2.33)
    {(a) Submitted-object validation};
  \node[box,text width=3.55cm] (a1) at (2.05,1.52)
    {Proposer-selected $\Gamma'$\\omits required $o^*$};
  \node[box,text width=3.15cm] (a2) at (6.35,1.52)
    {$\forall e\in\Gamma':$\\$\mathsf{Verify}(e)=1$};
  \node[key,text width=1.55cm] (a3) at (9.55,1.52) {ACCEPT};
  \node[note,text width=3.35cm] (a4) at (13.05,1.52)
    {$o^*$ is outside the\\verifier's quantifier};
  \draw[flow] (a1) -- node[above,font=\scriptsize]{submitted set only} (a2);
  \draw[flow] (a2) -- (a3);
  \draw[dashed,line width=0.55pt] (a3) -- (a4);
  \node[draw=black!35,fit=(a1)(a2)(a3)(a4),inner sep=5pt] {};

  \node[font=\footnotesize\bfseries,anchor=west] at (0.15,0.31)
    {(b) H-CRSPV exact-realization admission};
  \node[box,fill=black!7,text width=3.25cm] (b1) at (1.95,-0.52)
    {Authenticated $\rho$\\and the same $\Gamma'$};
  \node[box,fill=black!7,text width=3.35cm] (b2) at (6.05,-0.52)
    {$R=\Req(\rho)$\\$M=\Norm(\Gamma')$};
  \node[key,text width=3.75cm] (b3) at (10.45,-0.52)
    {Exact multiset comparison\\$(\pi^\rho_{\rm obl})_\#\mathsf{Occ}(M)\neq R$};
  \node[key,text width=2.75cm] (b4) at (14.45,-0.52)
    {REJECT BEFORE\\BACKEND DISPATCH};
  \draw[flow] (b1) -- (b2);
  \draw[flow] (b2) -- (b3);
  \draw[flow] (b3) -- (b4);
  \node[draw=black!35,fit=(b1)(b2)(b3)(b4),inner sep=5pt] {};
\end{tikzpicture}

%% file: figures/hcrspv_pipeline_tikz.tex
\begin{tikzpicture}[
  x=1cm,y=1cm,
  box/.style={draw=black!75, line width=0.55pt, fill=black!3,
              align=center, inner xsep=2.5pt, inner ysep=3.0pt,
              minimum height=1.02cm, font=\footnotesize},
  authority/.style={box, fill=black!12, font=\footnotesize\bfseries},
  proposer/.style={box, fill=black!7},
  validator/.style={box, fill=black!3},
  gate/.style={box, fill=black!9, font=\footnotesize\bfseries},
  terminal/.style={box, dashed, fill=white, font=\footnotesize\bfseries},
  panel/.style={draw=black!45, line width=0.45pt, rounded corners=0.5pt},
  flow/.style={-{Stealth[length=1.7mm,width=1.2mm]}, line width=0.55pt},
  fail/.style={flow,dashed},
  edgeword/.style={font=\footnotesize, fill=white, inner sep=1.2pt}
]
  \draw[panel] (-0.15,0.90) rectangle (16.65,4.55);
  \node[anchor=west,font=\footnotesize\bfseries] at (0.05,4.32)
    {(a) Exact obligation-to-object admission};

  \node[authority,text width=2.15cm] (required) at (1.25,3.20)
    {AUTHORITY\\required obligation\\occurrences $R$};
  \node[proposer,text width=2.15cm] (entries) at (4.05,3.20)
    {PROPOSER\\realization-entry\\occurrences};
  \node[validator,text width=2.15cm] (proposed) at (6.85,3.20)
    {VALIDATOR\\proposed groups\\quotient by $\kappa_\rho$};
  \node[gate,text width=2.15cm] (admissible) at (9.65,3.20)
    {ADMISSIBLE GROUPS\\registered-policy\\$\mathsf{GroupOK}_\rho$};
  \node[validator,text width=2.15cm] (objects) at (12.45,3.20)
    {CONCRETE OBJECTS\\one-to-one binding\\$\mu:G\xrightarrow{\sim}\mathcal O_\rho$};
  \node[gate,text width=2.15cm] (native) at (15.25,3.20)
    {NATIVE VERIFY\\proof or certificate\\for bound statement};

  \draw[flow] (required) -- node[edgeword] {one per occurrence} (entries);
  \draw[flow] (entries) -- node[edgeword] {\textsf{EntryComplete}} (proposed);
  \draw[flow] (proposed) -- node[edgeword] {policy check} (admissible);
  \draw[flow] (admissible) -- node[edgeword] {\textsf{ObjectBind}} (objects);
  \draw[flow] (objects) -- (native);

  \node[terminal,text width=5.2cm,minimum height=0.58cm] (reject) at (8.25,1.48)
    {\State{Rejected}: entry, grouping, binding, or native-evidence failure};
  \draw[fail] (entries.south) -- ++(0,-0.28) -| (reject.west);
  \draw[fail] (proposed.south) -- ++(0,-0.42) -| (reject.north west);
  \draw[fail] (objects.south) -- ++(0,-0.42) -| (reject.north east);
  \draw[fail] (native.south) -- ++(0,-0.28) -| (reject.east);

  \draw[panel] (-0.15,-2.25) rectangle (16.65,0.55);
  \node[anchor=west,font=\footnotesize\bfseries] at (0.05,0.30)
    {(b) Replicated release lifecycle};

  \node[validator,text width=2.65cm] (pending) at (2.15,-0.75)
    {\State{Pending}\\submitted, inactive};
  \node[gate,text width=3.05cm] (provisional) at (6.45,-0.75)
    {\State{Provisional}\\verified, challengeable, inactive};
  \node[authority,text width=2.65cm] (active) at (11.15,-0.75)
    {\State{Active}\\accepted for downstream use};
  \node[terminal,text width=2.65cm] (invalid) at (15.05,-0.75)
    {\State{Invalid}\\challenge upheld};

  \draw[flow] (pending) -- node[edgeword] {exact admission} (provisional);
  \draw[flow] (provisional) -- node[edgeword] {$h>d$; guarded finalizer} (active);
  \draw[fail] (provisional.south) -- ++(0,-0.47) -| node[edgeword,pos=0.76] {$h\leq d$} (invalid.south);
  \draw[fail] (pending.south) -- ++(0,-0.28) -| node[edgeword,pos=0.72] {upheld challenge} (invalid.south west);
  \node[terminal,text width=2.65cm,minimum height=0.58cm] (closed) at (2.15,-1.72)
    {\State{Rejected} or \State{Expired}};
  \draw[fail] (pending) -- (closed);
\end{tikzpicture}